\documentclass[twocolumn]{aastex631}

\newcommand{\Ni}{\ensuremath{^{56}\mathrm{Ni}}}

\newcommand{\Msun}{\ensuremath{\mathrm{M}_\odot}}

\newcommand{\RST}{\emph{Roman}}

\received{August 3, 2021}
\revised{November 30, 2021}
\accepted{December 7, 2021}

\shorttitle{High-redshift SN survey with Roman Space Telescope}
\shortauthors{Moriya, Quimby, \& Robertson}
\graphicspath{{./}{figures/}}

\begin{document}

\title{Discovering Supernovae at Epoch of Reionization with {\it Nancy Grace Roman Space Telescope}}

\author[0000-0003-1169-1954]{Takashi J. Moriya}
\affiliation{National Astronomical Observatory of Japan, National Institutes of Natural Sciences, 2-21-1 Osawa, Mitaka, Tokyo 181-8588, Japan}
\affiliation{School of Physics and Astronomy, Faculty of Science, Monash University, Clayton, Victoria 3800, Australia}

\author[0000-0001-9171-5236]{Robert M. Quimby}
\affiliation{Department of Astronomy / Mount Laguna Observatory, San Diego State University, 5500 Campanile Drive, San Diego, CA, 92812-1221, USA}
\affiliation{Kavli Institute for the Physics and Mathematics of the Universe (WPI), The University of Tokyo Institutes for Advanced Study, The University of Tokyo, 5-1-5 Kashiwanoha, Kashiwa, Chiba 277-8583, Japan}

\author[0000-0002-4271-0364]{Brant E. Robertson}
\affiliation{Department of Astronomy and Astrophysics, University of California, Santa Cruz, 1156 High Street, Santa Cruz, CA 96054, USA}



\begin{abstract}
Massive stars play critical roles for the reionization of the Universe. Individual massive stars at the reionization epoch ($z>6$) are too faint to observe and quantify their contributions to reionization. Some massive stars, however, explode as superluminous supernovae (SLSNe) or pair-instability supernovae (PISNe) that are luminous enough to observe even at $z>6$ and allow for the direct characterization of massive star properties at the reionization epoch. In addition, hypothetical long-sought-after PISNe are expected to be present preferentially at high redshifts, and their discovery will have a tremendous impact on our understanding of massive star evolution and the formation of stellar mass black holes. The near-infrared Wide Field Instrument on {\it Nancy Grace Roman Space Telescope} will excel at discovering such rare high-redshift supernovae. In this work, we investigate the best survey strategy to discover and identify SLSNe and PISNe at $z>6$ with \RST{}. We show that the combination of the F158 and F213 filters can clearly separate both SLSNe and PISNe at $z>6$ from  nearby supernovae through their colors and magnitudes. The limiting magnitudes are required to be 27.0~mag and 26.5~mag in the F158 and F213 filters, respectively, to identify supernovae at $z>6$. If we conduct a $10~\mathrm{deg^2}$ transient survey with these limiting magnitudes for 5 years with a cadence of one year, we expect to discover $22.5\pm2.8$ PISNe and $3.1\pm0.3$ SLSNe at $z>6$, depending on the cosmic star-formation history. The same survey is estimated to discover $76.1\pm8.2$ PISNe and $9.1\pm0.9$ SLSNe at $5<z<6$. Such a supernova survey requires the total observational time of approximately 525~hours in 5 years. The legacy data acquired with the survey will also be beneficial for many different science cases including the study of high-redshift galaxies.
\end{abstract}



\section{Introduction} \label{sec:introduction}
Massive stars play critical roles in shaping the Universe at its dawn. First-generation stars in the Universe are likely dominated by massive stars \citep[e.g.,][]{abel2002}. They quickly explode as supernovae (SNe) that provide chemical elements heavier than lithium in the Universe for the first time \citep[e.g.,][]{nomoto2013}. SNe also provide enormous kinetic energy to their surroundings, triggering the formation of subsequent generations of stars \citep[e.g.,][]{bromm2011}. Massive stars themselves, as well as their compact remnants such as the black holes and neutron stars they leave behind, play a critical role for the reionization of the Universe \citep[e.g.,][]{robertson2015}. The major sources of ionizing photons leading to the reionization of the Universe are not yet precisely known, and it is important to reveal the nature of massive stars in the early Universe to determine their exact contributions to reionization.

Massive stars in the early Universe evolve differently from those in the local Universe. The primary reason is that the massive stars in the early Universe have lower metallicities than the present-day massive stars. The lower metallicities make mass loss from massive stars less efficient \citep[e.g.,][]{vink2001,vink2005,mokiem2007,sander2020a,sander2020b}. Mass loss is a critical factor in determining the fates of massive stars \citep[e.g.,][]{smith2014}. It is also possible that the multiplicity of massive stars, another important factor in determining their fates, depends on metallicity \citep[e.g.,][]{machida2009}. The multiplicity of massive stars is important in stripping their hydrogen-rich envelopes during stellar binary evolution. The stripped stars become hotter and emit more radiation that can contribute to reionization \citep[e.g.,][]{stanway2016,gotberg2017,gotberg2020}. Stellar evolution in multiple systems strongly affects the final fates of massive stars as well \citep[e.g.,][]{podsiadlowski1992,nomoto1995,yoon2010,eldridge2013,laplace2021}. 

Unfortunately, it is difficult to observe directly individual stars in the reionization epoch ($z>6$), and to quantify their properties and contributions to cosmic reionization. Fortunately, however, some SNe from massive stars are luminous enough to discover even if they appear at $z>6$. Thus, it is possible to directly obtain information on massive stars at the reionization epoch by observing distant, luminous SNe. Superluminous SNe (SLSNe, \citealt{quimby2011}) are known populations of SNe that become luminous enough to probe the reionization epoch \citep[e.g.,][for a review]{gal-yam2019}. Although the exact nature of SLSNe is still unknown \citep[e.g.,][for a review]{moriya2018slsnreview}, hydrogen-poor (Type~I) SLSNe are known to originate from massive $(>8~\Msun)$ stars due to their presence in star-forming galaxies \citep[e.g.,][]{neill2011,leloudas2015,perley2016,schulze2020} and their estimated ejecta masses, which are large despite their lack of hydrogen \citep[e.g.,][]{nicholl2015,nicholl2019,blanchard2020}. In addition, hydrogen-poor SLSNe have been suggested to be a potential standardizable candle \citep[][]{inserra2014,inserra2021}. Therefore, high-redshift hydrogen-poor SLSNe may not only provide information on massive stars in the early Universe but also the distance to them. Hydrogen-rich (Type~II) SLSNe show the signatures of the interaction between the massive circumstellar matter (CSM) and SN ejecta that can explain their high luminosity \citep[e.g.,][]{smith2010}. Their progenitors have been considered to be massive stars because of the massive CSM ($\sim 10~\Msun$) required to explain their properties \citep[e.g.,][]{woosley2007,chevalier2011,chatzopoulos2012,moriya2013sn2006gy}, but the large opacity prevents us from investigating what kind of explosions occur within the massive CSM. Indeed, it was recently suggested that some hydrogen-rich SLSNe may be actually related to explosions of white dwarfs \citep[][]{jerkstrand2020}.

There also exist hypothetical SNe that are expected to be luminous enough even they appear at $z>6$: pair-instability SNe (PISNe). PISNe are predicted thermonuclear explosions of massive stars \citep[][]{barkat1967pisn,rakavy1967pisn}. Massive stars with helium core masses between $\simeq 65~\Msun$ and $\simeq 130~\Msun$ may explode as PISNe \citep[e.g.,][]{heger2002}. PISNe from massive cores produce a large amount of the radioactive \Ni\ that powers their light curves and they become extremely luminous \citep[e.g.,][]{kasen2011,dessart2013pisn,whalen2013pisnzero,kozyreva2014pisnprop,chatzopoulos2015pisnrot,gilmer2017pisn}. Although some SLSNe were previously suggested to be PISNe \citep[e.g.,][]{gal-yam2009sn2007bi}, most SLSNe are not likely PISNe and no conclusive PISN candidates have been observed (e.g., \citealt{dessart2012,jerkstrand2016pisnneb,moriya2019sn2007bi,mazzali2019}; see \citealt{terreran2017pisn} for a possible candidate). 
Massive helium cores above $\simeq 65~\Msun$ are required to power PISNe explosions. 
Mass loss prevents massive stars from sustaining the massive cores in the local Universe and PISNe are expected to occur exclusively at very low metallicity environments (e.g., \citealt{langer2007,yoshida2014}, but see also \citealt{georgy2017,higgins2021}). In addition, the massive stars that have enough mass to explode as PISNe are predicted to exist preferentially in the first generations of stars in the Universe \citep[e.g.,][]{hirano2015}. Discovering PISNe will not only be important to probe massive stars at high redshifts but also to confirm the predictions by the stellar evolution theory and the stellar formation theory in the early Universe. The peculiar chemical signatures of PISNe are not generally found (\citealt{debennassuti2017}, but see also \citealt{hartwig2018}), but some possible evidence has been reported \citep[][]{aoki2014}. 

Long-duration gamma-ray bursts (LGRBs) have been used to investigate massive star properties at high redshifts \citep[e.g.,][]{robertson2012}. LGRBs are very rare events and it is difficult to systematical discover high-redshift LGRBs. The uncertain LGRB properties such as jet angles also make them difficult to interpret as well \citep[e.g.,][]{goldstein2016}. Thus, having only LGRBs as a massive star probe is insufficient and information from high-redshift SNe from massive stars will improve our understanding of massive stars in the early Universe.

Deep optical transient surveys have discovered SNe up to $z\simeq 4$ \citep[][]{cooke2012,moriya2019shizuca,curtin2019shizuca}. 
In order to discover SNe at $z>6$, however, high-redshift transient surveys require deep near-infrared (NIR) observations \citep[][]{scannapieco2005,tanaka2013beyond6,pan2012,desouza2013,hartwig2018,moriya2019ultimate,wong2019ultimate,regos2020}. 
In addition, the capability to search for transients in a large area is essential because both SLSNe and PISNe are very rare. The total rate of SLSNe in the nearby Universe is around $200~\mathrm{Gpc^{-3}~yr^{-1}}$ \citep[][]{quimby2013}, while the local core-collapse SN rate is around $10^5~\mathrm{Gpc^{-3}~yr^{-1}}$ \citep[][]{madau2014}. Thus, only 0.1\% of exploding massive stars become SLSNe. The event rates of hypothetical PISNe are uncertain but similar rates to SLSNe have been predicted \citep[e.g.,][]{langer2007,dubuisson2020}.
Such rates are still consistent with the fact that we have never conclusively
observed PISNe \citep[][]{nicholl2013,moriya2021hsc}. 

Indeed, NIR wide-field surveys are at the frontier of the modern astronomy.
NIR wide-field surveys are best performed above the atmosphere to minimize background light and absorption bands, although a ground-based wide-field NIR facility with adoptive optics is currently under development at the Subaru Telescope\footnote{\url{https://ultimate.naoj.org/english/index.html}}.
The Euclid satellite will be launched in 2022, but the facility can only discover SLSNe up to $z\simeq 4$ given its survey design \citep[][]{inserra2018}. 
\RST{}, which is planned to be launched in the mid-2020's, will enable community-driven transient survey programs that can discover SNe at $z>6$.
\RST{} will likely have a time-domain high latitude survey to enable supernova cosmology \citep[][]{spergel2015,hounsell2018} and the discovery of other transients.
In this paper, we perform transient survey simulations for \RST{} to identify the best survey strategy to search for the SLSNe and PISNe at $z>6$.

The rest of this paper is organized as follows. We first show the PISN and SLSN models we use in our investigation in Section~\ref{sec:snmodels}. We then discuss the best filter combination for SN surveys to discover PISNe and SLSNe at $z>6$ in Section~\ref{sec:filter_selection}. We conduct mock SN survey simulations in Section~\ref{sec:simulations} to estimate the expected numbers of high-redshift SNe and determine the best survey cadence. 
We summarize our results in Section~\ref{sec:summary}. All magnitudes are in the AB system. We adopt the standard $\Lambda$CDM cosmology with $H_0=70~\mathrm{km~s^{-1}~Mpc^{-1}}$, $\Omega_\Lambda = 0.7$, and $\Omega_M =0.3$ when necessary.

\section{Supernova models}\label{sec:snmodels}
We first introduce the SN models used in this study. We consider PISNe and SLSNe as potential high-redshift
targets owing to their high luminosity. 

\subsection{PISN}
We take PISN models calculated by \citet{kasen2011}. The PISN explosions are calculated by the stellar evolution code \texttt{Kepler} and the explosion properties such as explosion energies and $^{56}\mathrm{Ni}$ mass are self-consistently obtained \citep[][]{heger2002}. \citet{kasen2011} perform numerical multi-frequency radiation transport calculations of the PISN models and the time evolution of the rest-frame spectral energy distribution (SED) is obtained. In order to compute the high-redshift PISN light curves in the observer frame for a filter at a redshift, we take the synthetic rest-frame SEDs, shift them to the observer frame assuming the given redshift, and convolve with the filter response function.

\citet{kasen2011} study PISNe from red supergiant (RSG), blue supergiant (BSG), and Wolf-Rayet (WR) progenitors. The RSG progenitors have $10^{-4}~\mathrm{Z_\odot}$ and the BSG and WR progenitors have zero-matellicity. Recent low-metallicity single massive star evolution models consistently show that PISNe explode as RSGs when they evolve as single stars \citep[e.g.,][]{yoon2012,moriya2015pisn}. PISNe from WR stars can exist especially when we take binary stellar evolution into account. For example, chemically homogeneous evolution of massive close binary stars can make massive WR stars that explode as PISNe \citep[e.g.,][]{marchant2016}. Thus, we take the PISN models from RSG and WR progenitor stars in this study.

PISN explosion properties strongly depend on their progenitor mass \citep[][]{heger2002}. More massive progenitors become more luminous PISNe because a larger amount of the radioactive $^{56}\mathrm{Ni}$ is produced. Among the PISN models from RSGs, we take the three most massive models; R200 (200~\Msun\ RSG progenitor), R225 (225~\Msun\ RSG progenitor), and R250 (250~\Msun\ RSG progenitor). For the PISNe from WR stars, we consider the three most massive models; He130 (130~\Msun\ WR progenitor), He120 (120~\Msun\ WR progenitor), and He110 (110~\Msun\ WR progenitor).

\begin{figure}
\epsscale{1.2}
\plotone{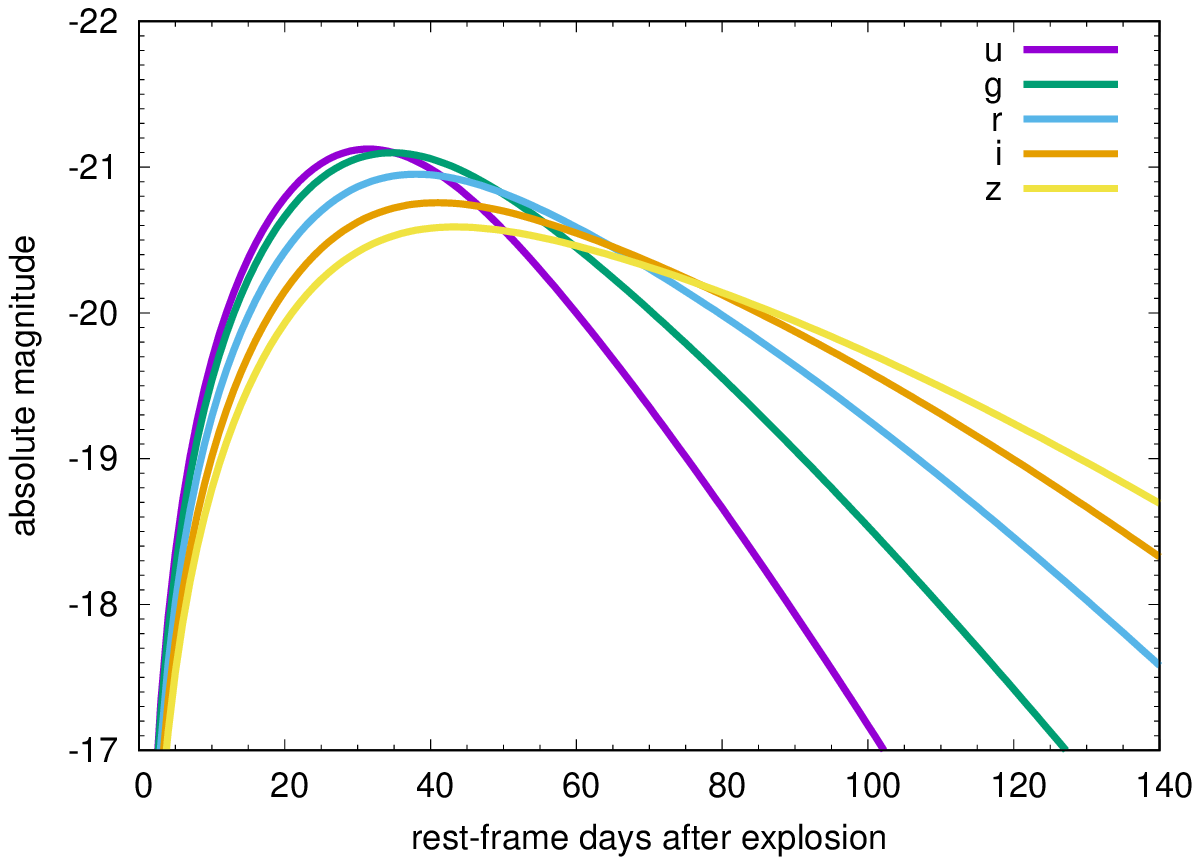}
\plotone{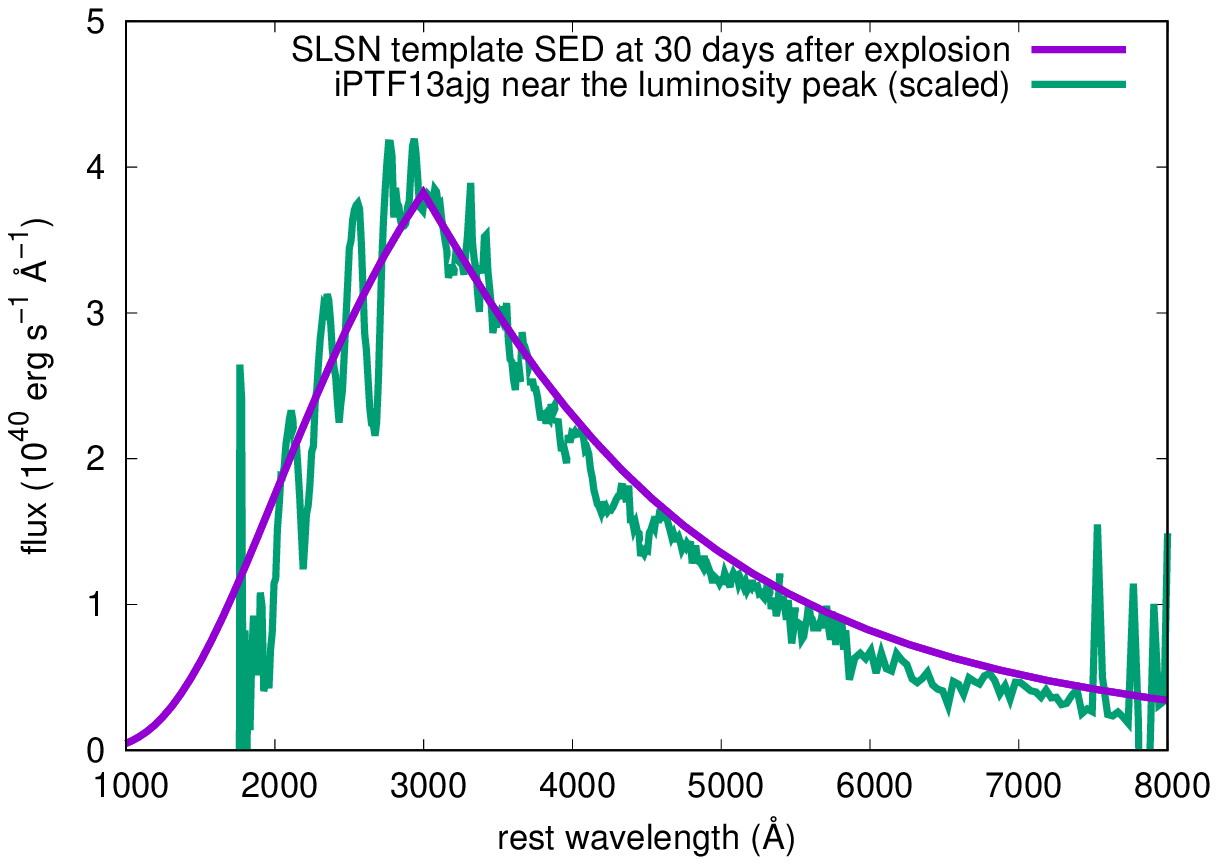}
\caption{
\textit{Top:} Rest-frame optical light curves of our SLSN template. \textit{Bottom:} Comparison between the SED of the SLSN template at 30~days after explosion and the spectral shape of SLSN iPTF13ajg near the luminosity peak.
\label{fig:slsntemplate}}
\end{figure}

\subsection{SLSN}\label{sec:slsntemplate}
From the two SLSN subclasses (Type~II with hydrogen features and Type~I without them), we consider Type~I SLSNe in this study. A large fraction of SLSNe reported in literature are Type~I SLSNe and we have good knowledge of their luminosity function and timescales. Meanwhile, only several Type~II SLSNe have been published and general knowledge on their photometric and spectroscopic properties is still poor. 
Thus, we consider only Type~I SLSNe in this study. 
The event rate of Type~II SLSNe is estimated to be higher than that of Type~I SLSNe \citep[][]{quimby2013}. Therefore, the following discovery number estimates with only Type~I SLSNe should be conservative. We refer to Type~I SLSNe as SLSN~I or simply as SLSNe in the rest of this paper.

The SLSN light-curve template we use here is constructed in the same method as in \citet{moriya2021hsc}, and as originally presented by \citet{prajs2017}. 
Briefly, we assume that the luminosity source of SLSNe is the spin-down of strongly magnetized neutron stars (magnetars) as proposed by \citet{kasen2010,woosley2010}. Regardless of the actual nature of SLSNe, the magnetar model is shown to explain the photometric properties of SLSNe around their peak luminosity well \citep[e.g.,][]{inserra2013,liu2017,nicholl2017,hsu2021}. Assuming the spin-down energy deposition at the center of SN ejecta, we obtain the luminosity evolution based on the semi-analytic formula developed by \citet{arnett1982}. The magnetar model requires the specification of the initial magnetar rotational period ($P$), the magnetar magnetic field ($B$), and the diffusion timescale in the SN ejecta ($\tau_d$).

The model rest-frame SEDs are obtained by assuming the blackbody function at the photosphere. The location of photosphere is determined by assuming the constant opacity of $0.1~\mathrm{cm^2~g^{-1}}$. Because SLSN spectra below 3000~\AA\ do not follow the blackbody function, we truncate the SEDs below 3000~\AA\ to match the spectral shape of the well-observed SLSN iPTF13ajg spectrum at around the luminosity peak (\citealt{vreeswijk2014}, Fig.~\ref{fig:slsntemplate}). Some SLSNe are known to have ultraviolet luminosity excess \citep[e.g.,][]{nicholl2017gaia16apd}, but such an ultraviolet-bright SLSN would be easier to discover at high redshifts. We refer to \citet{moriya2021hsc} for further details on the SLSN template generation.

The average peak magnitude of SLSNe at 4000~\AA\ are estimated to be $-21.1~\mathrm{mag}$ \citep{lunnan2018pansample}. Their typical rise time is around 30~days \citep{lunnan2018pansample,decia2018,angus2019}. Thus, we set the model parameters to have the $u$ band (3543~\AA) peak magnitude of $-21.1~\mathrm{mag}$ and rise time of 30~days. The adopted model parameters are $P=2.05~\mathrm{ms}$, $B=3\times 10^{13}~\mathrm{G}$, and $\tau_d = 21~\mathrm{days}$. We show the SLSN template in Fig.~\ref{fig:slsntemplate}. 
The peak magnitudes of SLSNe are estimated to vary with the standard deviation of 0.7~mag \citep[][]{lunnan2018pansample}. Thus, in the following survey simulations, we randomly shift the template magnitudes assuming the Gaussian probability distribution with the standard deviation of 0.7~mag. Similarly, the rise time of SLSNe shows diversity. We scaled the templates so that the rise times randomly change following a Gaussian probability distribution with a standard deviation of 10~days, with a minimum of 10~days to prevent unrealistically small rise times.
To summarize, our survey simulations include SLSNe with rise times of $30\pm 10~\mathrm{days}$ and peak $u$ band magnitudes of $-21.1\pm 0.7~\mathrm{mag}$ with Gaussian scatter.

\section{Filter selection}\label{sec:filter_selection}
\begin{figure}
\epsscale{1.2}
\plotone{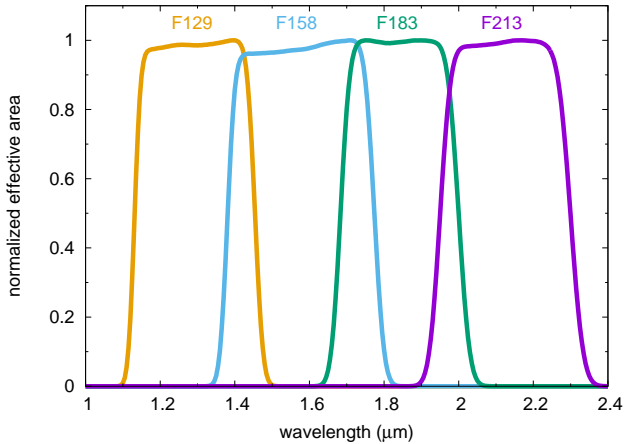}
\caption{
Normalized effective areas of the four \RST{} filters we use in this study.
\label{fig:filters}}
\end{figure}

In this section we discuss the best filter combination to discover and identify SNe at $z>6$.
We adopt the latest information on the filters and limiting point-source sensitivities released on 21 January 2021\footnote{\url{https://roman.gsfc.nasa.gov/science/Roman_Reference_Information.html}}. We consider the four \RST{} NIR filters (F129, F158, F184, and F213) with effective areas presented in Fig.~\ref{fig:filters}. Among the four filters we consider, we use simulations to determine the best two-filter combination for the high-redshift SN survey. We assume a two-filter survey because color information is essential for selecting high-redshift SN candidates. Although having more than two filters is ideal, it requires more observational time.

\begin{figure*}
 \begin{center}
  \includegraphics[width=1.05\columnwidth]{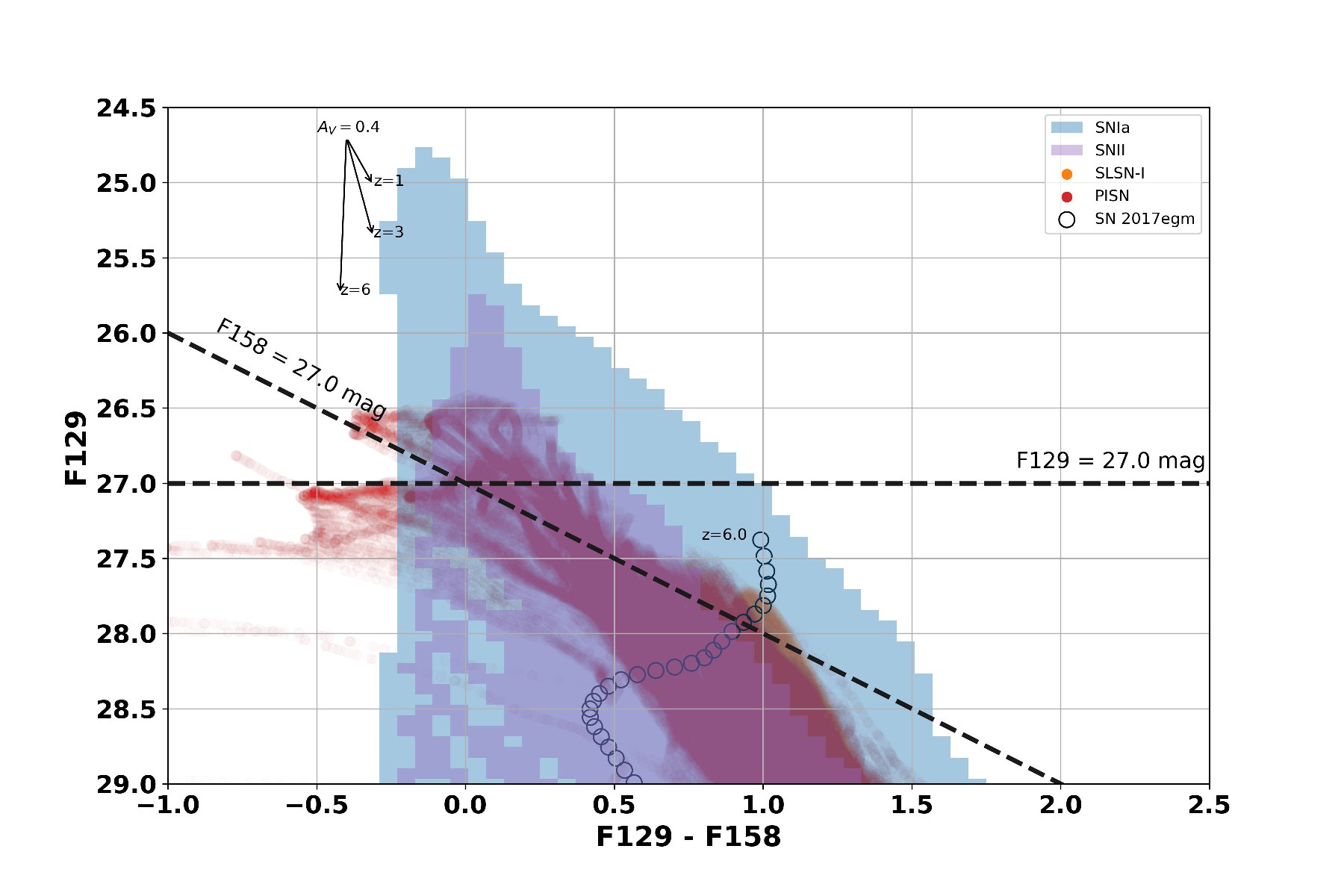}   
  \includegraphics[width=1.05\columnwidth]{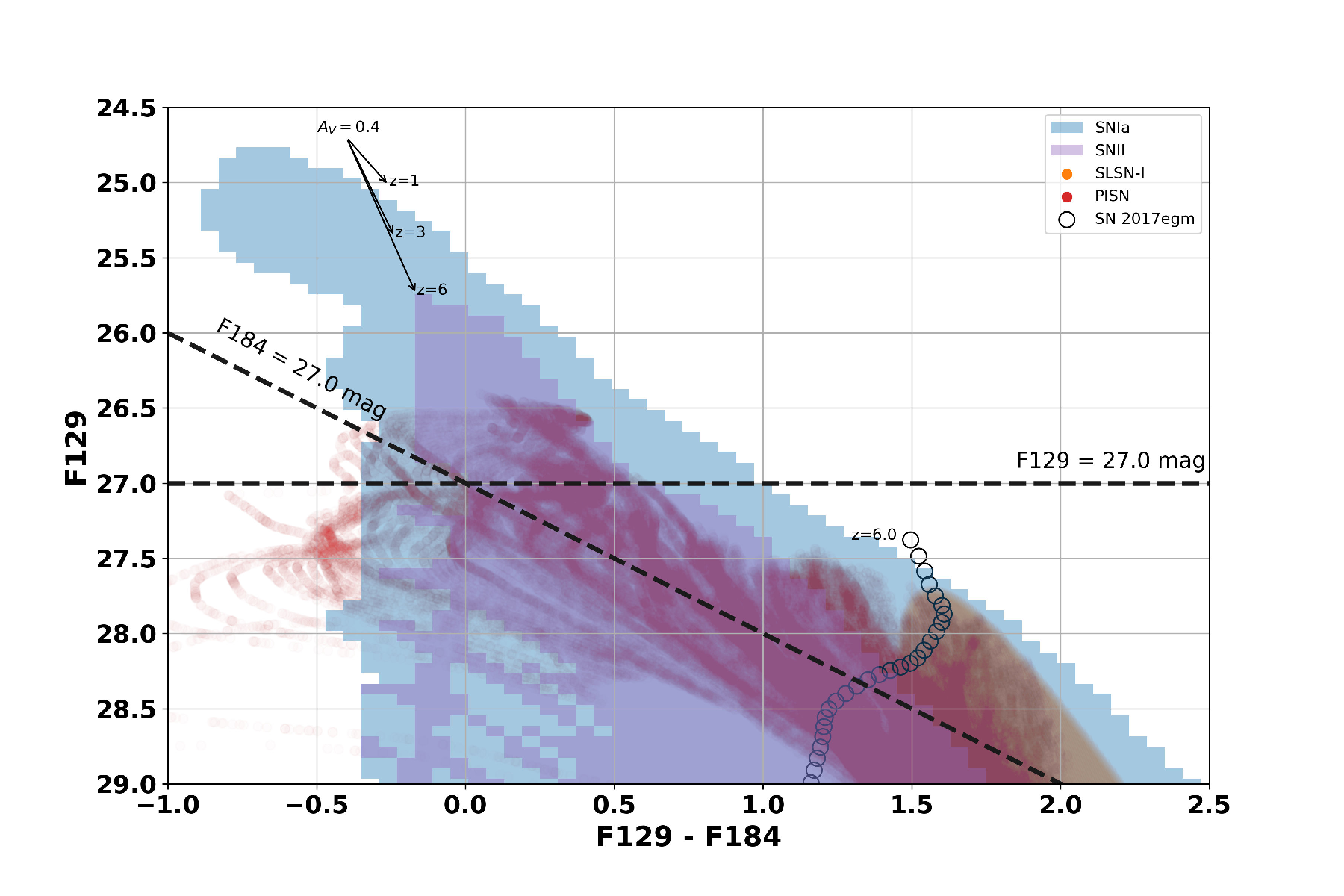}    
  \includegraphics[width=1.05\columnwidth]{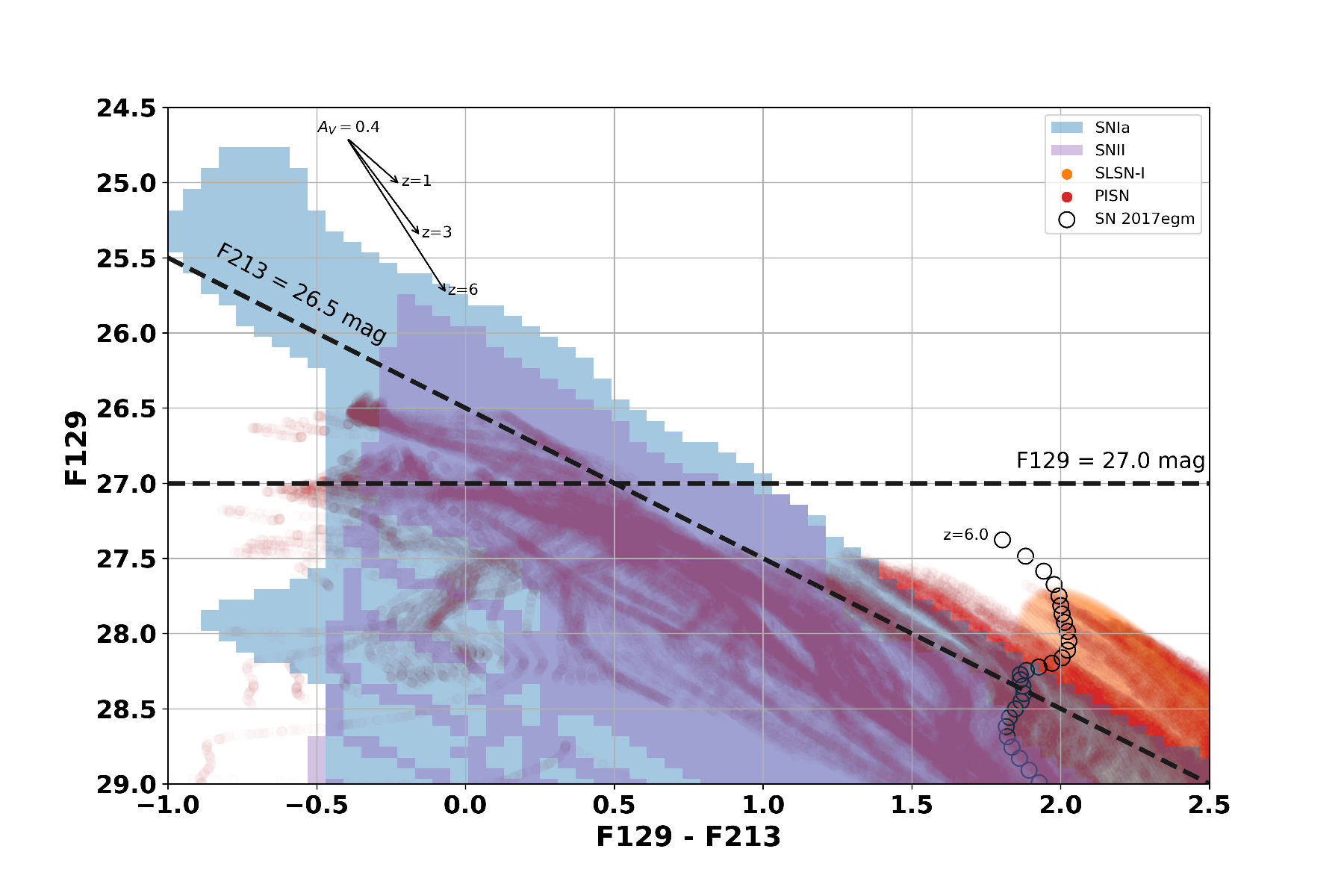}   
  \includegraphics[width=1.05\columnwidth]{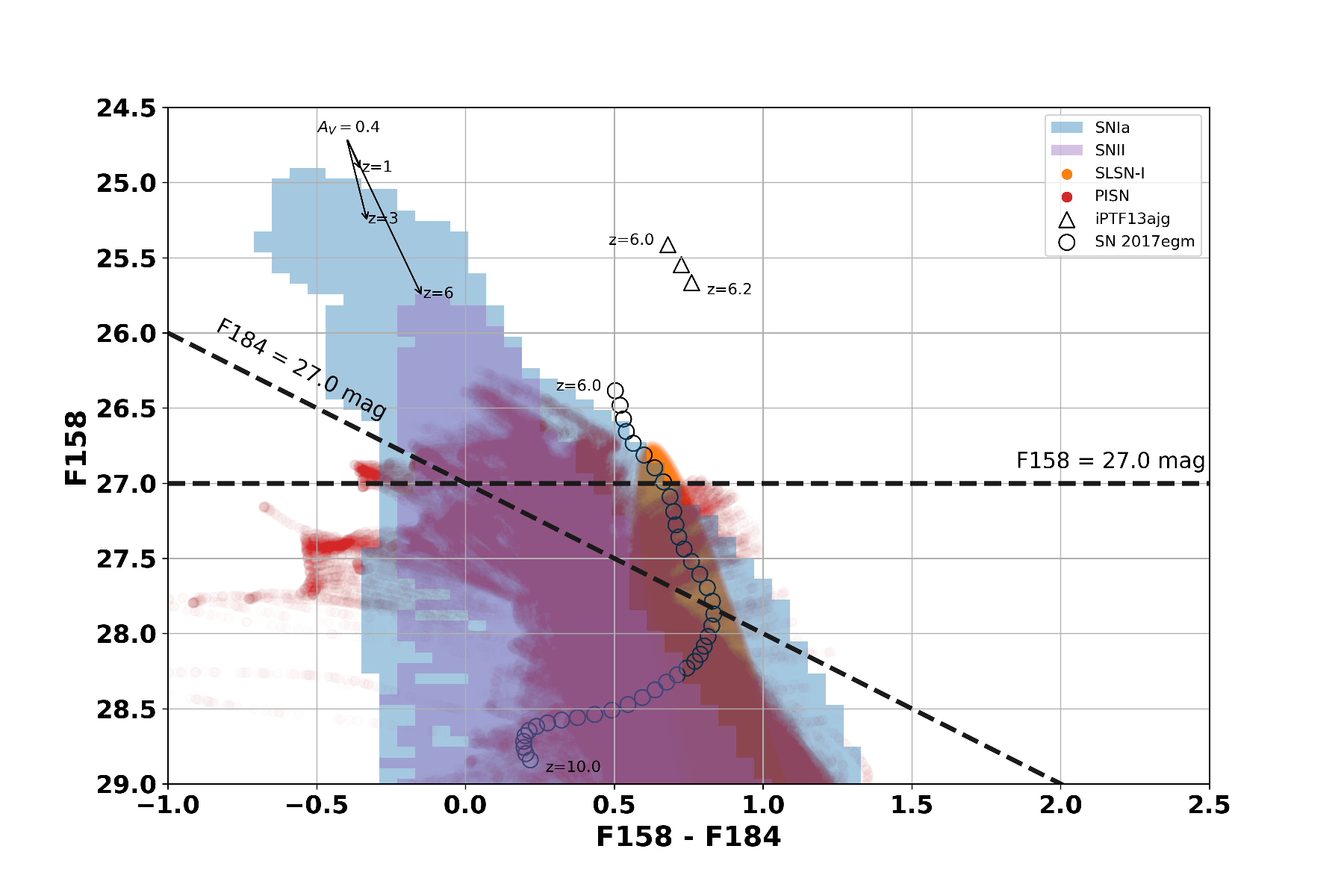}    
  \includegraphics[width=1.05\columnwidth]{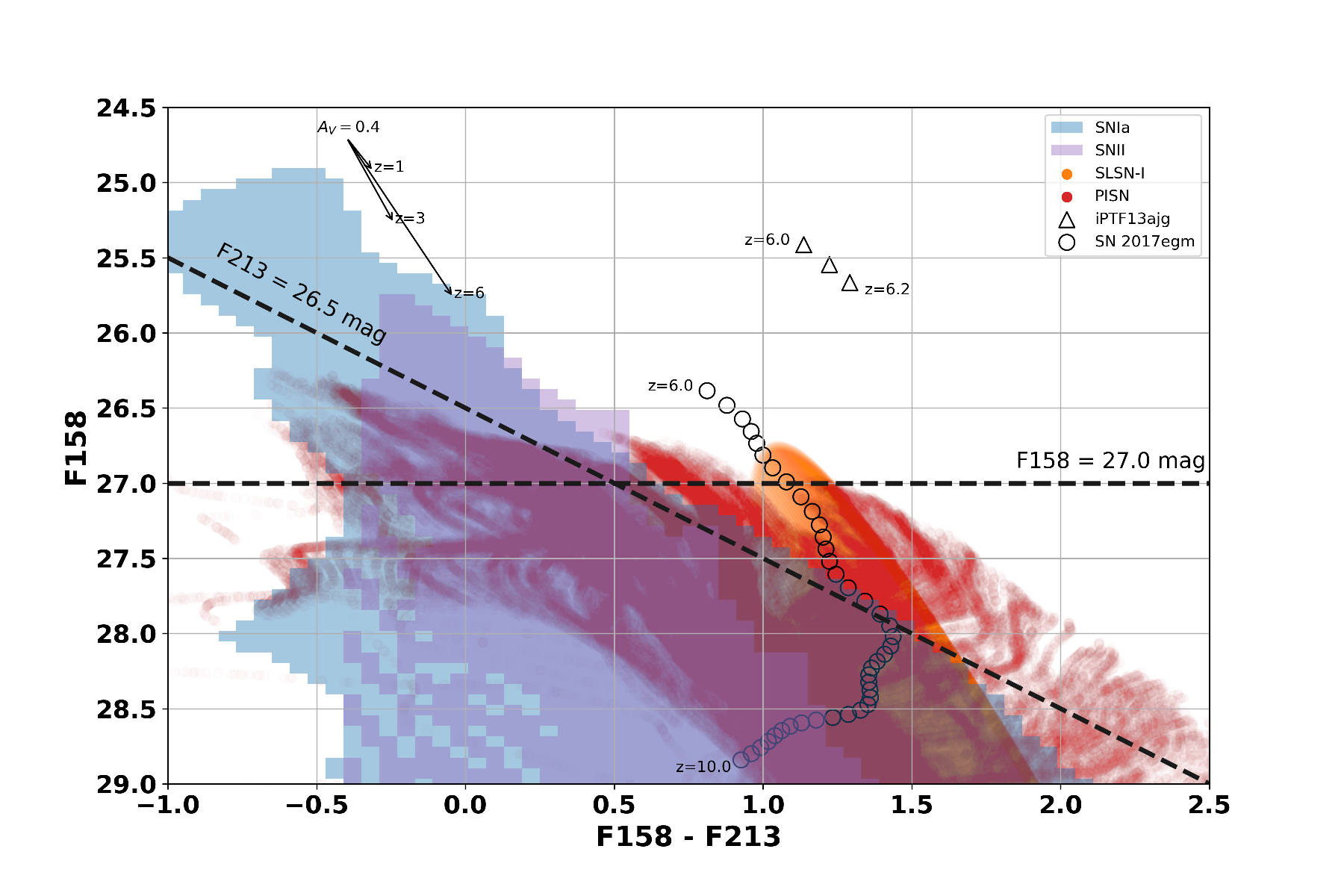}  
  \includegraphics[width=1.05\columnwidth]{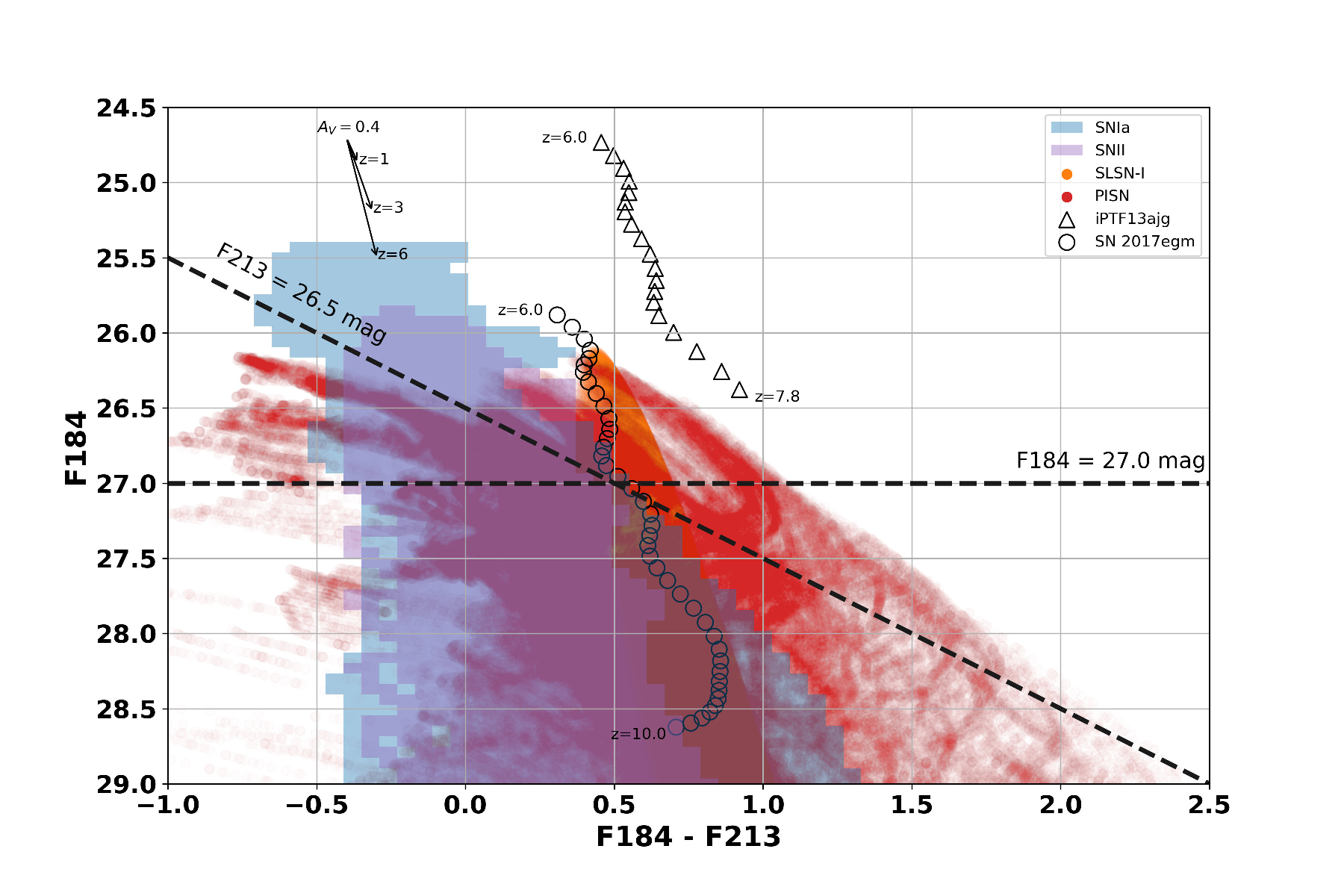}  
 \end{center}
\caption{
Color-magnitude diagrams for six possible \RST{} filter combinations. ``SNIa'' and ``SNII'' show the regions occupied by Type~Ia and Type~II SNe at $z>1.5$, respectively, with an extinction of $A_V=0,1,$ or 2~mag. ``SLSN-I'' and ``PISN'' show the regions occupied by our SLSN and luminous PISN models, respectively, at $6<z<10$. In addition, the expected locations based on the observed spectra at around the peak luminosity of two SLSNe, iPTF13ajg \citep[][]{vreeswijk2014} and SN~2017egm \citep[][]{bose2018,nicholl2017sn2017egm}, are shown. No extinction is applied in the SLSN and PISN models, but the extinction vectors are shown at the top left of each panel. We note that the host galaxy extinction at $z>6$ is likely small \citep[e.g.,][]{bouwens2021,dunlop2013,dunlop2012}.
}\label{fig:cmd}
\end{figure*}

Fig.~\ref{fig:cmd} shows the color-magnitude diagrams (CMDs) of SNe at all phases. The figure includes Type~Ia and Type~II SNe at $z>1.5$. These SNe will be as bright as PISNe and SLSNe at $z>6$ and we need to efficiently exclude them to identify SNe at $z>6$. The colors and magnitudes of Type~Ia SNe are estimated based on the Type~Ia SN spectral template of \citet{hsiao2007}. The template luminosity is set to $-19.3~\mathrm{mag}$ at the peak in the SDSS \textit{r} band, which is the average of the standard Type~Ia SNe \citep[][]{yasuda2010}. For Type~II SNe, we use Nugent's template\footnote{\url{https://c3.lbl.gov/nugent/nugent_templates.html}}. The peak luminosity is set to $-18.5~\mathrm{mag}$ at the peak in the SDSS \textit{r} band, brighter than the average peak luminosity of Type~II SNe \citep[][]{richardson2014}. We assume the brighter peak luminosity because brighter Type~II SNe are more likely to be discovered in the flux-limited surveys we assume in this paper. Type~Ia and Type~II SNe shown in Fig.~\ref{fig:cmd} include extinctions of $A_V =0, 1,$ and 2~mag. 

Examining the CMDs, we find that the combinations of (F129, F158), (F129, F184), and (F158, F184) do not allow us to clearly distinguish high-redshift PISNe and SLSNe from Type~Ia and Type~II SNe. In other words, F213 is essential in distinguishing SNe at $z>6$ from nearby ones. In this paper, we choose to have a filter combination that can identify both PISN and SLSN candidates, because having information on both PISNe and SLSNe, which come from different mass ranges in massive stars, is beneficial to constrain massive star properties at $z>6$. The $\mathrm{F184}-\mathrm{F213}$ color of high-redshift SLSNe has only 0.2~mag difference at most from those of Type~Ia and Type~II SNe, while the $\mathrm{F129}-\mathrm{F213}$ and $\mathrm{F158}-\mathrm{F213}$ colors can have the difference of more than 0.5~mag. Thus, the combinations of (F129, F213) and (F158, F213) are preferred to clearly distinguish both PISNe and SLSNe from Type~Ia and Type~II SNe.
In order to distinguish them, F129 needs to reach at least 28.0~mag while F158 only needs to reach 27.0~mag at least.
The sensitivities of F129 and F158 for \RST{} are roughly the same (the $5\sigma$ limit of 28.0~mag with one-hour exposure). Therefore, we conclude that the best filter combination for high-redshift PISN and SLSN identification with \RST{} is F158 and F213. The $\mathrm{F158}-\mathrm{F213}$ CDM shows that transient surveys with limiting magnitudes of 27.0~mag or deeper in F158 and 26.5~mag or deeper in F213 can separate Type~Ia and Type~II SNe at $z>1.5$ from PISNe and SLSNe at $z>6$.
For example, if we adopt $\mathrm{F158}=27.0~\mathrm{mag}$ and $\mathrm{F213}=26.5~\mathrm{mag}$ limits, transients with a detection in F213 and a non-detection in F158 can be identified as high-redshift PISN and SLSN candidates (Fig.~\ref{fig:cmd}).
Simultaneous ground-based optical transient surveys with Subaru and/or Vera Rubin Observatory can also help with efficient filtering of lower redshift Type~Ia and Type~II SNe.

\begin{figure*}
 \begin{center}
  \includegraphics[width=1.05\columnwidth]{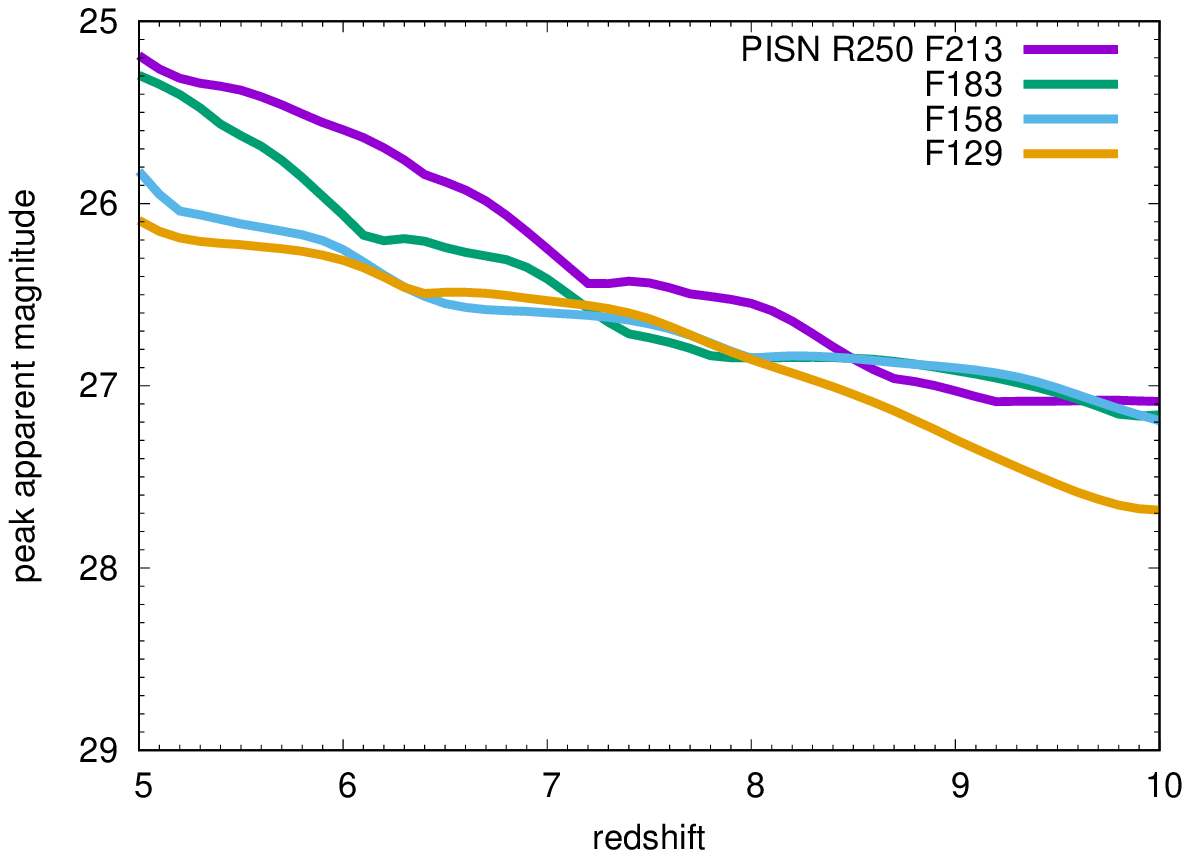}   
  \includegraphics[width=1.05\columnwidth]{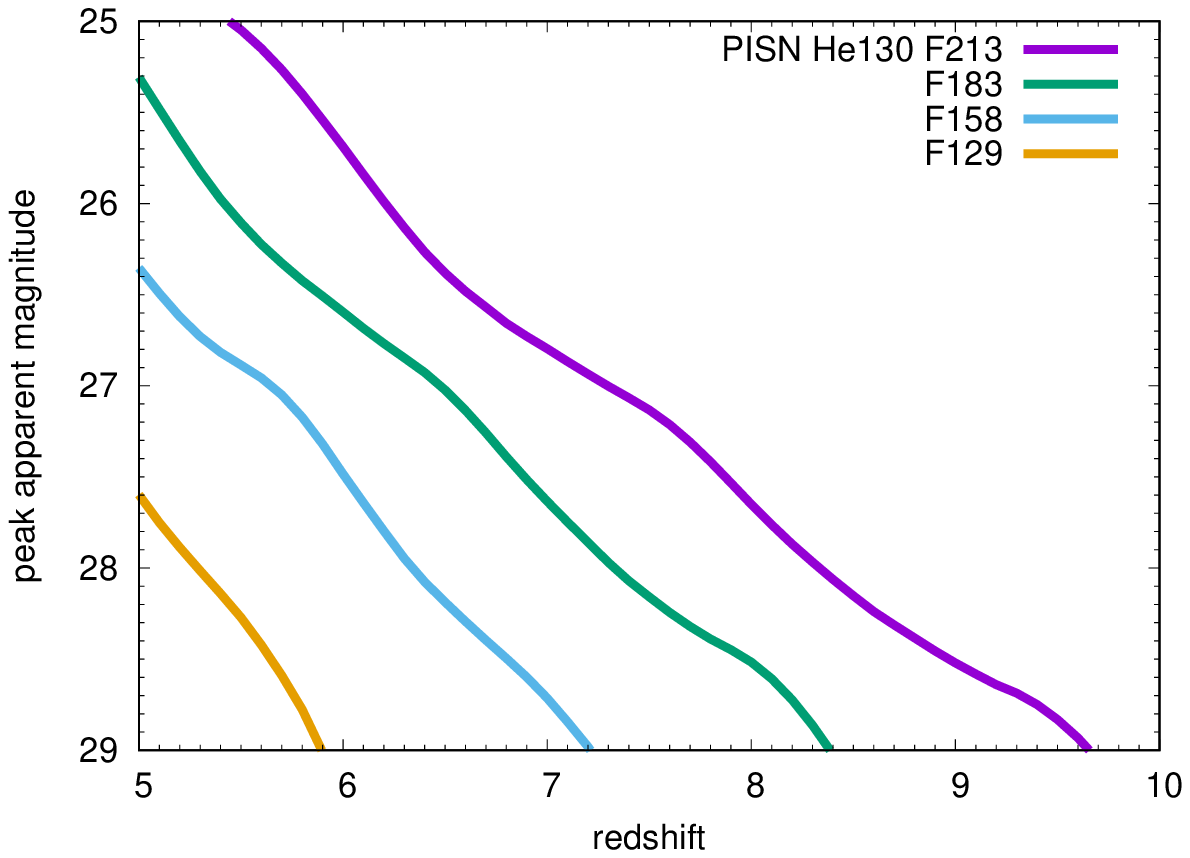}   
  \includegraphics[width=1.05\columnwidth]{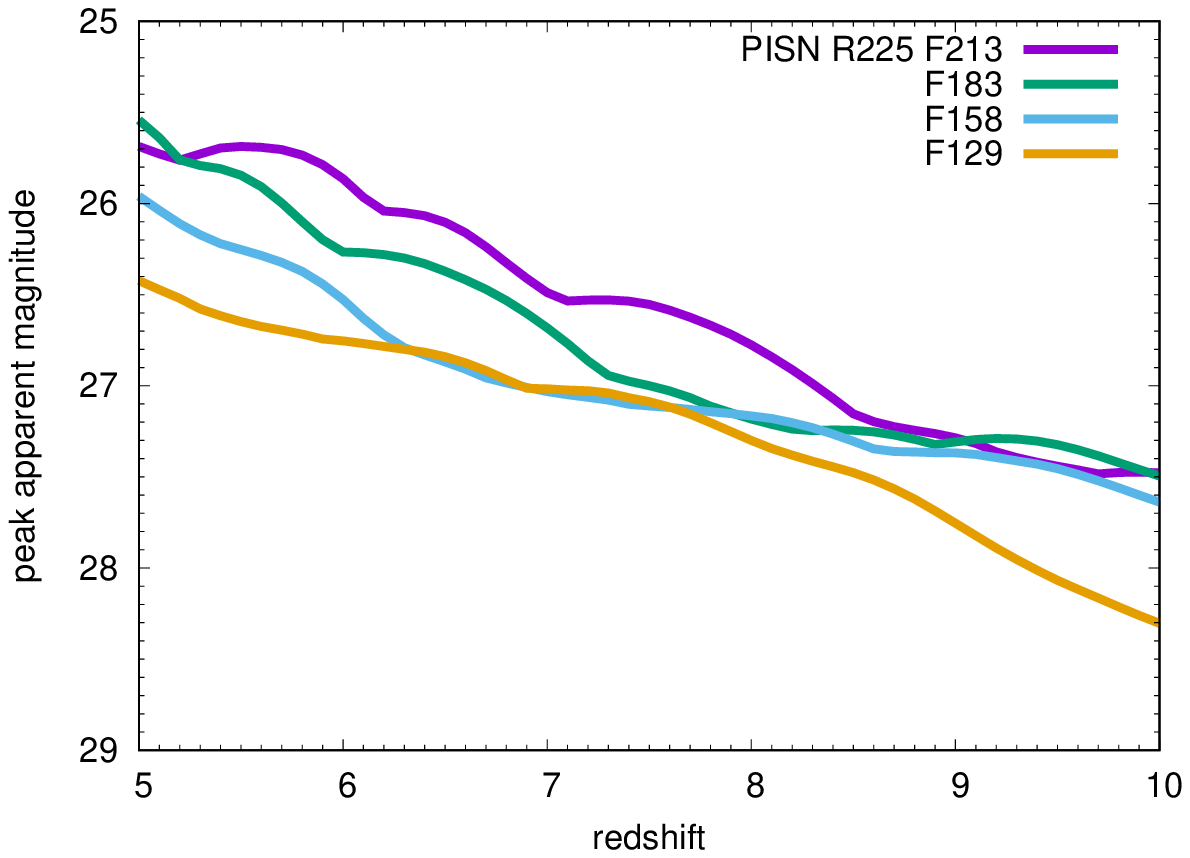} 
  \includegraphics[width=1.05\columnwidth]{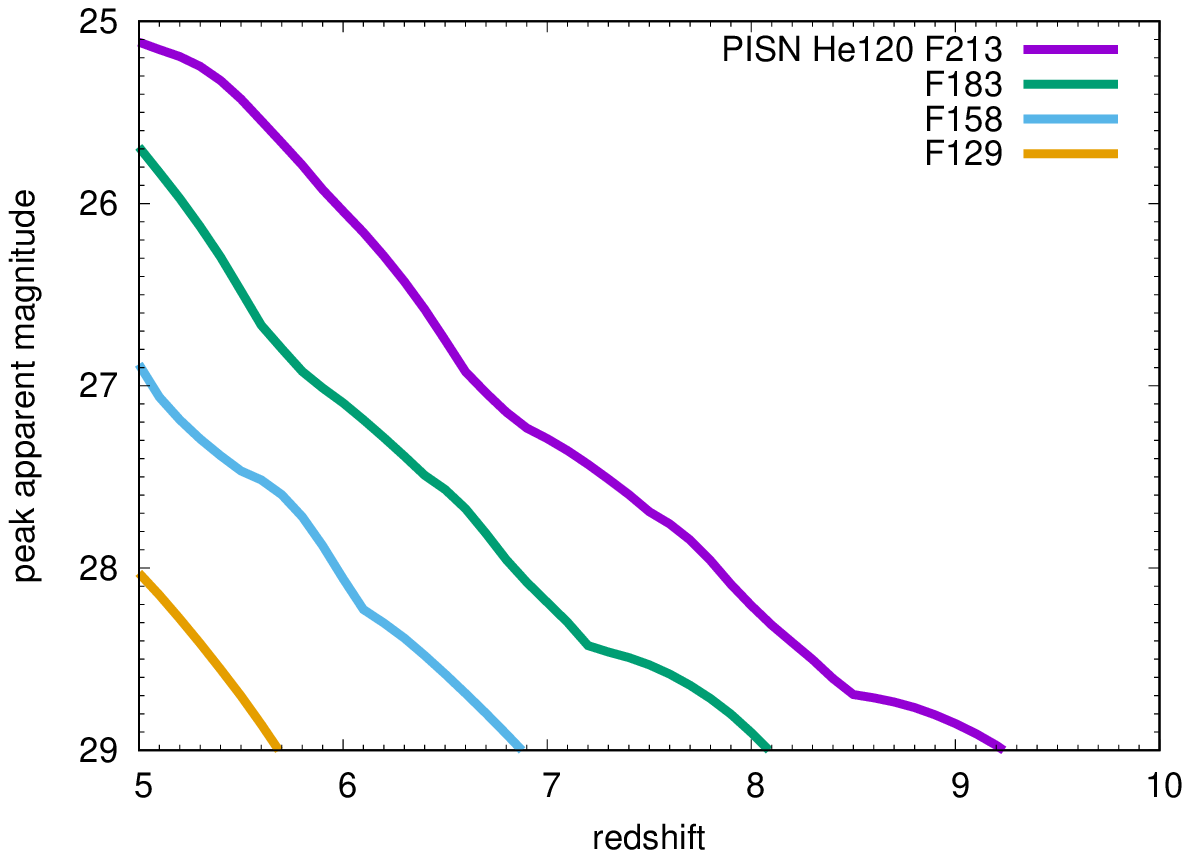}   
  \includegraphics[width=1.05\columnwidth]{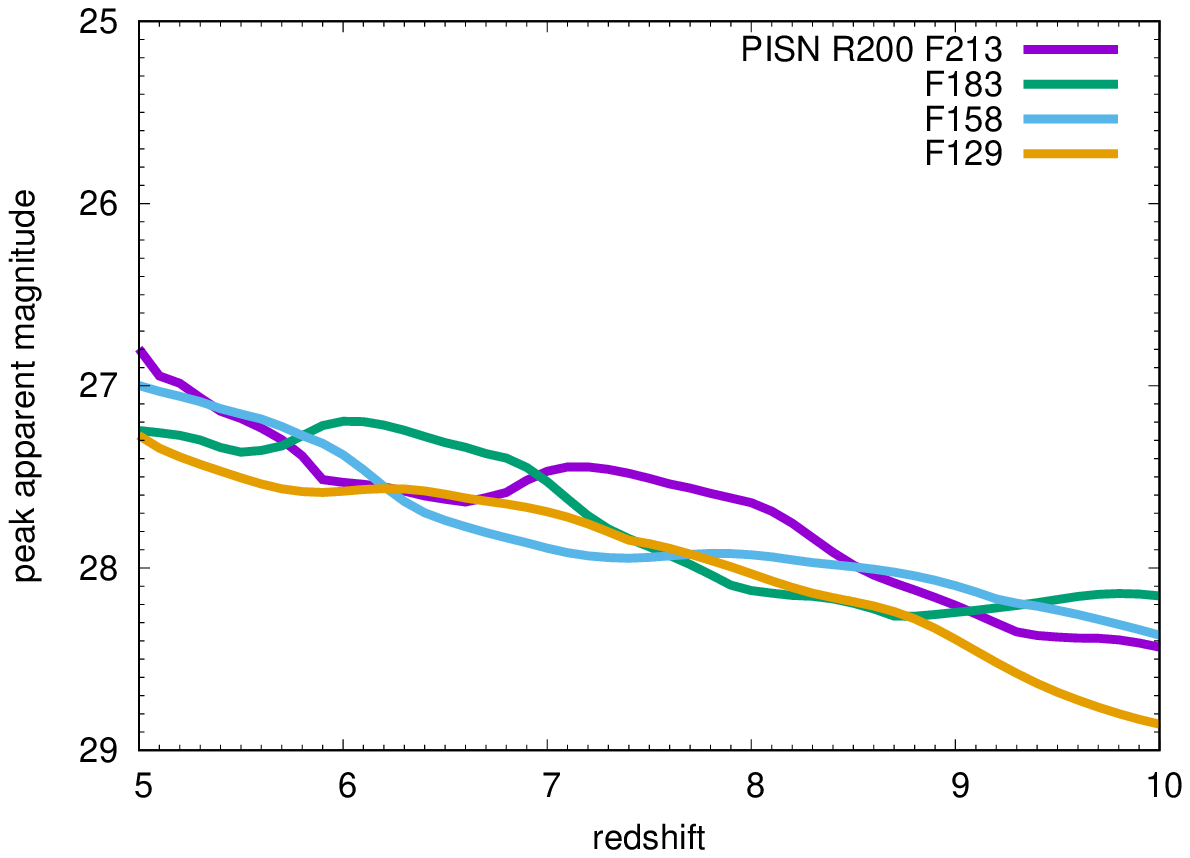} 
  \includegraphics[width=1.05\columnwidth]{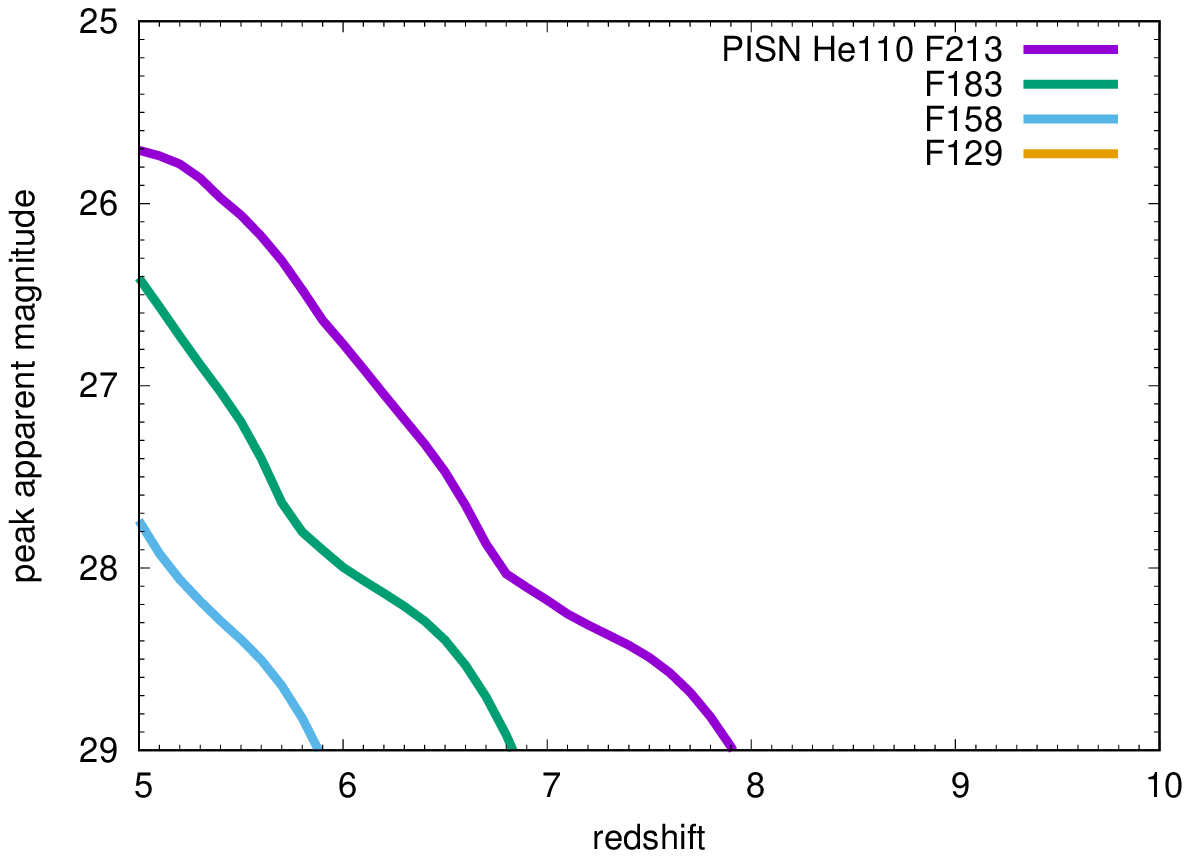} 
 \end{center}
\caption{
Peak apparent magnitudes of the PISN models from the RSG (left) and WR star (right) progenitors at each redshift.
}\label{fig:peakmag_pisn}
\end{figure*}

\begin{figure}
\epsscale{1.2}
\plotone{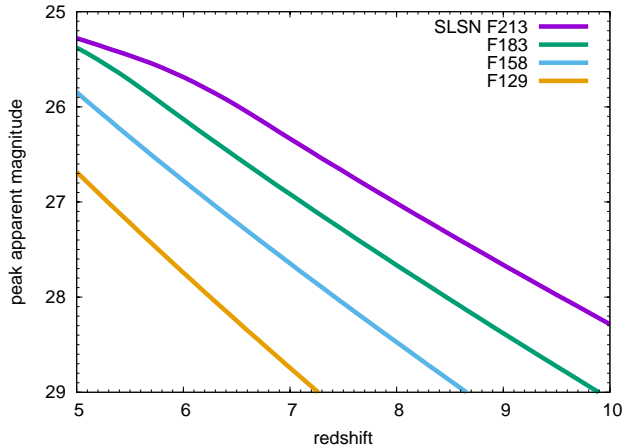}
\caption{
Peak apparent magnitudes of the SLSN template at each redshift. Note that we randomly modify the magnitudes of the SLSN template in the survey simulations as described in Section~\ref{sec:simulations}.
\label{fig:peakmag_slsn}}
\end{figure}

\begin{figure*}
 \begin{center}
  \includegraphics[width=1.05\columnwidth]{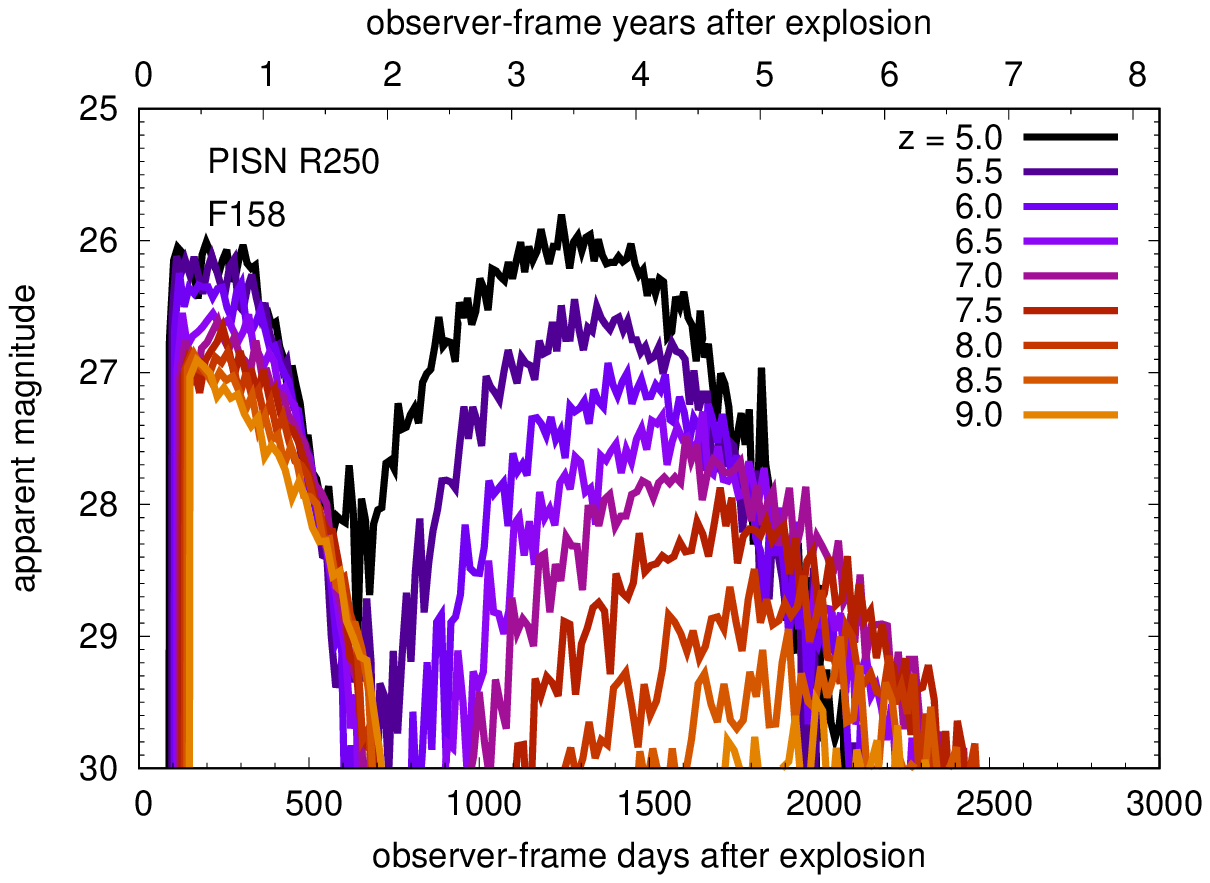}   
  \includegraphics[width=1.05\columnwidth]{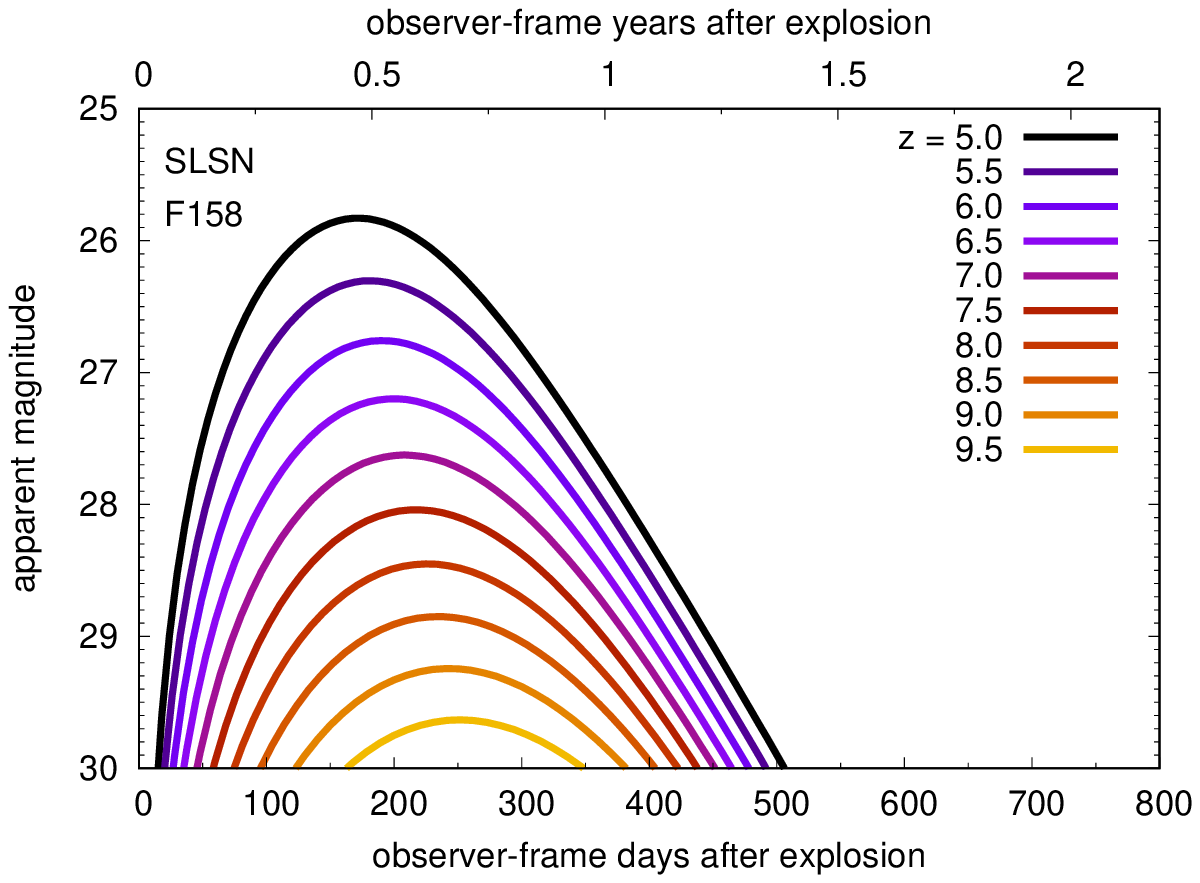}   
  \includegraphics[width=1.05\columnwidth]{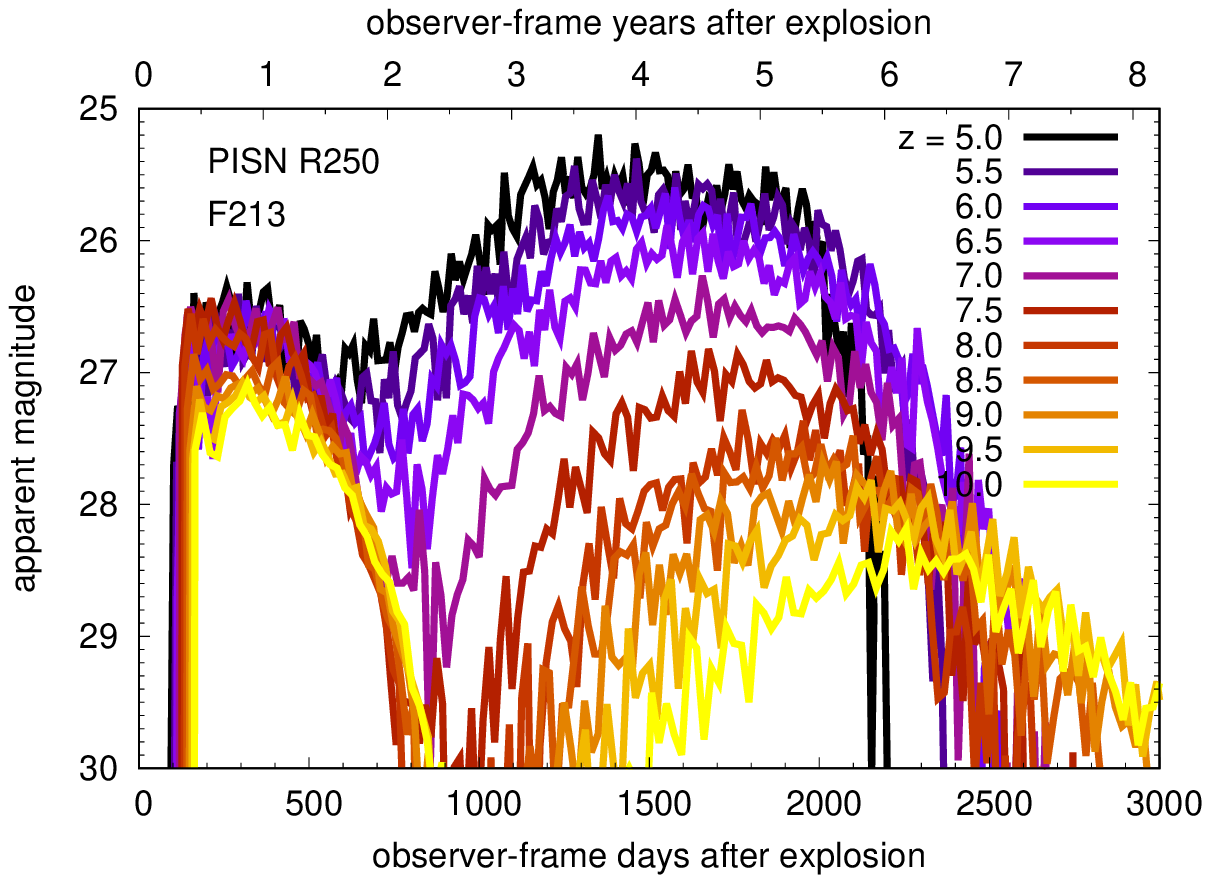} 
  \includegraphics[width=1.05\columnwidth]{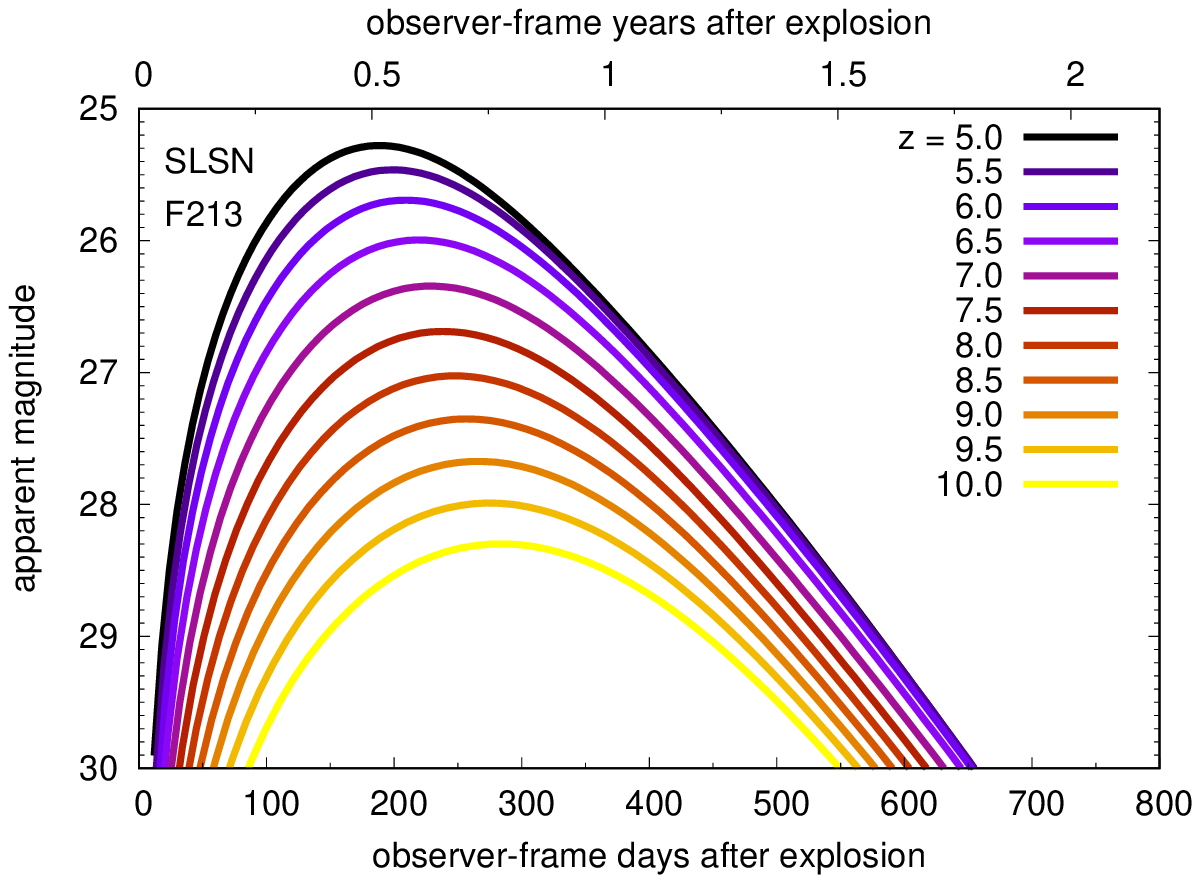}   
 \end{center}
\caption{
Observer-frame light curves of the R250 PISN model (left) and SLSN template (right) in the \RST{} F158 (top) and F213 (bottom) filters. We randomly modify the light-curve duration and magnitudes of the SLSN template in the survey simulations as described in Section~\ref{sec:snmodels}.
}\label{fig:lightcurves}
\end{figure*}

Fig.~\ref{fig:peakmag_pisn} shows the peak apparent magnitudes of luminous PISN models from the RSG and WR progenitors. The R250, R225, He130, and He120 PISN models become brighter than the $\mathrm{F158} = 27.0~\mathrm{mag}$ and/or $\mathrm{F213} = 26.5~\mathrm{mag}$ limits even if they appear at $z>6$. Fig.~\ref{fig:peakmag_slsn} shows that SLSNe up to $z\simeq7.2$ become bright enough to discover with the same limits. The light curves of the R250 PISN and SLSN at $z>6$ in the F158 and F213 filters are presented in Fig.~\ref{fig:lightcurves}.

\section{Survey simulations}\label{sec:simulations}
We have shown that a \RST{} survey to discover and identify SNe at $z>6$ would best be conducted by using the F158 and F213 filters with limiting magnitudes of 27.0~mag and 26.5~mag, respectively. In this section, we conduct SN survey simulations to estimate the expected numbers of SNe at $z>6$ that can be discovered with \RST{} and examine efficient survey strategy considerations such as cadence. We first describe the assumptions in our survey simulations in the next section.

\subsection{Survey simulation setups}\label{sec:setups}
\subsubsection{Survey area, duration, and cadence}\label{sec:area_duration_cadence}
We assume a survey area of $10~\mathrm{deg^2}$. The possibility of making deep imaging fields of around $10~\mathrm{deg^2}$ has been considered by the \RST{} project \citep{spergel2015}. The deepest layer of \RST{} Type~Ia SN surveys previously considered is also $5-10~\mathrm{deg^2}$ \citep[][]{hounsell2018}. Thus, $10~\mathrm{deg^2}$ is a reasonable area for a deep transient survey with \RST{}.
The expected numbers of high-redshift SN discovery are proportional to the survey area and our results can be scaled to match other possible survey designs. The \RST{} Wide Field Instrument has an effective field-of-view of $0.281~\mathrm{deg^2}$ and requires 36 pointings to cover a $10~\mathrm{deg^2}$ area \citep{hounsell2018}.

We assume a survey duration of 5~years, which is the nominal operation period of \RST{}. The survey area needs to be observed repeatedly to discover high-redshift SNe during the survey period. High-redshift SNe experience time dilation and high-cadence observations are not necessarily required to discover them. We use 0.5~yr, 1.0~yr, and 1.5~yr cadences in our survey simulations and discuss the best cadence based on the simulation results in the following sections. 

\subsubsection{Filters and sensitivities}\label{sec:filters_limits}
Transient surveys with the \RST{} F158 and F213 filters with limiting magnitudes of $\mathrm{F158}=27.0~\mathrm{mag}$ and $\mathrm{F213}=26.5~\mathrm{mag}$ are found to be the best combination of filters and depths in Section~\ref{sec:filter_selection}. The $5\sigma$ limiting point-source sensitivities of F158 and F213 in one-hour exposure are 28.0~mag and 26.2~mag, respectively. Thus, 571~s is required to reach 27.0~mag in F158 and 6257~s is required to reach 26.5~mag in F213. We assume that reference images that are 0.5~mag deeper than the survey depth are obtained at the beginning of the transient survey. The required times to reach 27.5~mag in F158 and 27.0~mag in F213 with \RST{}
are 1434~s and 15715~s, respectively.

\subsubsection{SN rates}
\begin{figure}
\epsscale{1.2}
\plotone{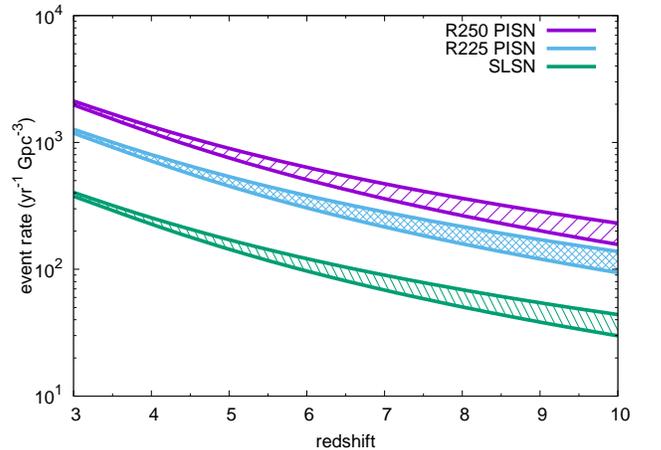}
\caption{
SN event rates in our survey simulations. They are based on the SFR density estimated by \citet{robertson2015}. The hatched regions show the possible ranges of the SN rates that originate from uncertainties in the SFR density.
\label{fig:snrate}}
\end{figure}

Another essential ingredient in SN survey simulations is the SN event rates. Fig.~\ref{fig:snrate} presents the SN rates assumed in our survey simulations and we describe them in this section. SNe in our simulations are randomly generated based on the assumed SN rates.

In order to estimate the SN rates, we first need to assume the star-formation rate (SFR) history. We adopt the cosmic SFR density history estimated by \citet{robertson2015}. We assume no delay time between the star-formation and SN explosion, because both PISNe and SLSNe are explosions of massive stars that occur shortly ($\sim 10~\mathrm{Myr}$) after the birth of the progenitors.
The SFR density history is an important assumption in the SN rate estimates. For example, the cosmic SFR density we adopt is much lower than those assumed in a previous study by \citet{pan2012} and our number estimates are, therefore, more conservative.

Given the SFR density, we need to assume the fractions of stars that explode as PISNe and SLSNe to obtain their event rates. For PISNe, we estimate the fraction based on the Salpeter initial mass function (IMF). The R250 and R225 PISN models, which become bright enough to discover at $z>6$ (Fig.~\ref{fig:peakmag_pisn}), are the explosions of 250~\Msun\ and 225~\Msun\ stars at the zero-age main sequence, respectively. We assume that the properties of PISNe from stars between $237.5~\Msun$ and $300~\Msun$ (the mass interval $\Delta M=62.5~\Msun$) are represented by the R250 model and those from stars between $212.5~\Msun$ and 237.5~\Msun\ ($\Delta M = 25~\Msun$) are represented by the R225 model. This assumption is necessary because the only progenitor masses of PISN models provided by \citet{kasen2011} are 200~\Msun, 225~\Msun, and 250~\Msun.
We simply assume that the closest-mass available PISN model represents the PISN properties of the given mass. The maximum mass that causes the PISN explosion is set at 300~\Msun. The mass range of forming stars is assumed to be from 0.1~\Msun\ to 500~\Msun. 
We do not include PISNe from WR stars in our survey simulations because of their uncertain event rates. As we do not consider PISNe from WR stars, the expected numbers of PISN discovery presented here should be conservative. The R250 and R225 PISN event rates obtained under these assumptions are presented in Fig.~\ref{fig:snrate}. The R250 PISN rate is estimated to be slightly higher than the R225 PISN rate despite of the Salpeter IMF because of the larger mass interval for the R250 model ($\Delta M=62.5~\Msun$) compared with the R225 model ($\Delta M = 25~\Msun$).

For SLSNe, we obtain the event rates by setting their progenitor fraction in the nearby Universe. \citet{quimby2013} find that the SLSN~I rate at $z\simeq 0.2$ is $31^{+74}_{-25}~\mathrm{yr^{-1}~Gpc^{-3}}$. A more recent study by \citet{frohmaier2021} found a similar SLSN~I rate of $35^{+25}_{-13}~\mathrm{yr^{-1}~Gpc^{-3}}$ at $z\lesssim 0.2$. Because SLSNe preferentially appear in low metallicity environments \citep[e.g.,][]{chen2017slsnhostmetallicity,schulze2018,wiseman2020}, it is likely that the SLSN fraction in the early Universe is larger than what we observe in the nearby Universe. Thus, we adopt the SLSN fraction that provides the nearby SLSN rate of $100~\mathrm{yr^{-1}~Gpc^{-3}}$ in our survey simulations (Fig.~\ref{fig:snrate}).

\subsubsection{Required observational time}\label{sec:required_time}

The total required observational time depends on the cadence, the exposure time ($t_\mathrm{exp}$), the number of pointings, and the overhead for each exposure. The required exposure time to reach a certain limiting magnitude is provided in Section~\ref{sec:filters_limits}. 36 pointings are required to cover $10~\mathrm{deg^2}$ in each epoch. The overhead time $t_\mathrm{oh}$ is uncertain. The overhead time of 42~s for each exposure was assumed in early SN survey simulations with \RST{} \citep[][]{spergel2013}, but \citet{hounsell2018} argue that 42~s is significantly underestimated. To make conservative estimates, we assume the overhead time of 100~s in this paper.

We assume two-filter transient surveys in this paper. The required observational time at each epoch ($t_\mathrm{ep}$) is, therefore,
\begin{equation}
    t_\mathrm{ep} = 36 (t_\mathrm{exp,1} + N_1 t_\mathrm{oh}) + 36 (t_\mathrm{exp,2} + N_2 t_\mathrm{oh}), \label{eq:texp}
\end{equation}
where $t_\mathrm{exp,1}$ and $t_\mathrm{exp,2}$ are the exposure time to reach the assumed sensitivity for the two filters used in the transient survey. The factors $N_1$ and $N_2$ account for the number of exposures required to reach the limiting magnitude in one epoch. We assume $N_1=N_2=1$ in this paper. Dithered exposures are not required for the SN survey because the \RST{} Wide Field Instrument features H4RG-10 detectors and cosmic rays can be removed without dithering by non-destructive reads.

\begin{deluxetable*}{ccccccccccc}
\tablecaption{
Expected numbers of the high-redshift PISN and SLSN discovery from our survey simulations. Average numbers within the assumed SN rates are presented. The associated uncertainties indicate the range of the SN discovery numbers within the uncertain SN rates. The survey area is $10~\mathrm{deg^2}$ and the survey duration is 5~years. $t_\mathrm{total}$ is the total required observational time for the survey. \label{tab:discoverynumbers}
}
\tablewidth{0pt}
\tablehead{
\multicolumn{2}{c}{Limit} & & & & & & & & \\ 
\colhead{F158} & \colhead{F213} & \colhead{Cadence} & \colhead{$t_\mathrm{total}$} & \colhead{$z>5.0$} & \colhead{$z>6.0$} & \colhead{$z>6.2$} & \colhead{$z>6.4$} & \colhead{$z>6.6$} & \colhead{$z>6.8$} & \colhead{$z>7.0$} 
}
\startdata
\multicolumn{11}{c}{PISN} \\
27.0~mag & 26.5~mag & 0.5~yr & 877~hr & $78.9\pm8.5$ & $24.2\pm3.3$ & $17.7\pm2.3$ & $12.1\pm1.6$ & $7.3\pm1.0$ & $3.5\pm0.5$ & $1.2\pm0.1$ \\ 
27.0~mag & 26.5~mag & 1.0~yr & 525~hr & $76.1\pm8.2$ & $22.5\pm2.8$ & $16.0\pm2.1$ & $10.5\pm1.4$ & $5.9\pm0.7$ & $2.1\pm0.2$ & $0.62\pm0.08$  \\
27.0~mag & 26.5~mag & 1.5~yr & 385~hr & $64.1\pm6.9$ & $18.4\pm2.2$ & $13.0\pm1.7$ & $8.3\pm1.1$ & $4.5\pm0.5$ & $1.5\pm0.1$ & $0.40\pm0.06$ \\ 
\hline
\multicolumn{11}{c}{SLSN} \\
27.0~mag & 26.5~mag & 0.5~yr & 877~hr & $12.0\pm1.2$ & $4.4\pm0.5$ & $3.4\pm0.4$ & $2.7\pm0.3$ & $2.0\pm0.2$ & $1.5\pm0.1$ & $1.1\pm0.1$ \\
27.0~mag & 26.5~mag & 1.0~yr & 525~hr &  $9.1\pm0.9$ & $3.1\pm0.3$ & $2.4\pm0.2$ & $1.8\pm0.2$ & $1.3\pm0.1$ & $1.0\pm0.1$ & $0.76\pm0.09$ \\ 
27.0~mag & 26.5~mag & 1.5~yr & 385~hr &  $5.9\pm0.6$ & $1.9\pm0.2$ & $1.5\pm0.2$ & $1.1\pm0.1$ & $0.8\pm0.1$ & $0.6\pm0.1$ & $0.45\pm0.09$  
\enddata
\end{deluxetable*}

\begin{figure}
\epsscale{1.2}
\plotone{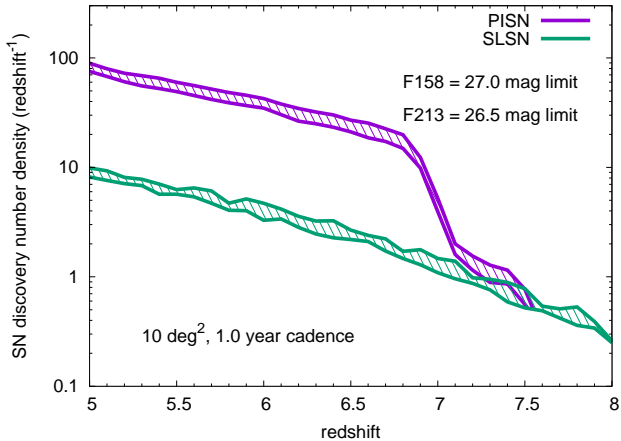}
\caption{
High-redshift SN discovery number density distributions estimated by our survey simulations. The hatched regions indicate the possible range originating from SN rate uncertainties (Fig.~\ref{fig:snrate}).
\label{fig:redshiftdistribution}}
\end{figure}

\begin{figure}
\epsscale{1.2}
\plotone{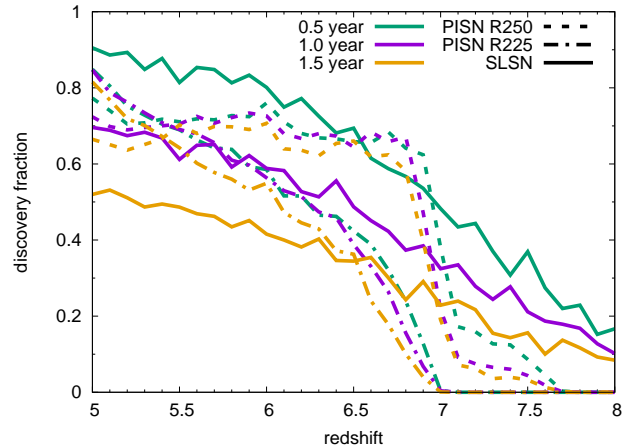}
\caption{
Discovery fractions of SNe in the surveys with the 0.5~year, 1.0~year, and 1.5~year cadences. The fractions are obtained by dividing the discovery numbers by all the SNe exploded during the survey.
\label{fig:efficiency}}
\end{figure}

To search for transients, we need to obtain reference images at the beginning of the survey so that we can find newly appearing objects in the sky. The reference images need to be deeper than the survey images. In our survey simulations, the reference images are assumed to be taken at the beginning of the 5-year survey. When we simulate the 5-year transient survey with the 1-year cadence, for example, the reference images are assumed to be obtained on the 1st day of the survey. Then, the subsequent survey images are taken with the one-year interval up to 5~years. We search for SNe that are fainter than the limiting magnitudes of the reference images and become brighter than the limiting magnitudes of the survey images. The required time to take the reference images is
\begin{equation}
    t_\mathrm{ref} = 36 (t_\mathrm{exp,1,\mathrm{ref}} + N_{1,\mathrm{ref}} t_\mathrm{oh}) + 36 (t_\mathrm{exp,2,\mathrm{ref}} + N_{2,\mathrm{ref}} t_\mathrm{oh}), \label{eq:tref}
\end{equation}
where the subscript ``ref'' indicates the required values to take the reference images that are deeper than the survey images. We assume that the limiting magnitudes of the reference images are 0.5~mag deeper than those during the survey images in our simulations. We set $N_{1,\mathrm{ref}}=N_{2,\mathrm{ref}}=1$ as before.

The total required observational time $t_\mathrm{total}$ for the entire survey is estimated as
\begin{equation}
    t_\mathrm{total} = N_\mathrm{ep} t_\mathrm{ep} + t_\mathrm{ref},
\end{equation}
where $N_\mathrm{ep}$ is the number of epochs in the surveys.

\subsection{Survey simulation results}\label{sec:results}
We show the results of the survey simulations obtained under the assumptions explained so far. Only PISNe and SLSNe at $z>6$ that become brighter than 26.5~mag in F213 can be distinguished from the closer Type~Ia and Type~II SNe with their colors and magnitudes (Section~\ref{sec:filter_selection}). Thus, PISNe and SLSNe that become brighter than 26.5~mag in F213 at least one time are regarded as a discovery in our survey simulations. 
We assume that the reference images are taken at the beginning of the survey period. We count the number of SNe that are fainter than the limits of the reference images and become brighter than the assumed limit during the survey.
We repeat the survey simulations with the same assumptions for 1000 times and show the average number of discovered SNe. No extinction is assumed in the survey simulations. The host galaxy extinction is likely small in the star-forming galaxies at $z\gtrsim 6$ likely to host the SNe of interest \citep[e.g.,][]{bouwens2021,dunlop2013,dunlop2012}.
We note that it is still possible that PISNe and SLSNe are affected by an extinction in a local star-forming environment even if the host galaxy extinction is generally small.
Galactic extinction can be avoided by judiciously choosing the survey field.

Table~\ref{tab:discoverynumbers} summarizes the expected numbers of SN discovery from our survey simulations. The redshift distributions of the discovered SNe are presented in Fig.~\ref{fig:redshiftdistribution}. We find that both PISNe and SLSNe at $z>6$ can be discovered with \RST{} $10~\mathrm{deg^2}$ deep transient surveys with limiting magnitudes of $\mathrm{F158} = 27.0~\mathrm{mag}$ and $\mathrm{F213} = 26.5~\mathrm{mag}$. The 1.0~year cadence survey is, for example, expected to discover $22.5\pm2.8$ PISNe at $z>6$ and $3.1\pm0.3$ SLSNe at $z>6$, depending on the uncertain cosmic SFR density.
The expected SN discovery numbers are proportional to the cosmic SFR density. The uncertainty in the cosmic SFR density directly affects the expected number of discovery.

In our survey simulations, we can count the numbers of SNe that exploded but are missed by the surveys and obtain the fractions of SNe that are discovered by the simulated surveys. Fig.~\ref{fig:efficiency} shows the discovery fractions. These fractions are obtained by dividing the numbers of the discovery in the surveys by the number of the SNe exploded during the survey period. The figure shows that the discovery fraction generally decreases with the longer cadences, especially in SLSNe that have a shorter timescale compared to PISNe. While the shorter cadence surveys are better to discover more SNe and to obtain better event rate constraints, they require more time. The 0.5~year and 1.0~year cadence surveys allow us to discover higher redshift SLSNe than the 1.5~year cadence survey. In addition, conducting the transient survey in the same season every year with the 1.0~year cadence would maximize the effective survey area because the search area will be decreased due to pointing constraints and the geometry of the detector array, which is not symmetric to rotation, by observing the same field in a different season. Overall, we suggest that the 1.0~year cadence survey is a reasonable choice. The same survey can discover $76.1\pm8.2$ PISNe and $9.1\pm0.9$ SLSNe at $5<z<6$ (Table~\ref{tab:discoverynumbers}).

A further benefit of shorter cadence surveys is that high-redshift SNe are more likely to be discovered in multiple epochs. With multiple epoch detection, we can use light-curve information to select high-redshift SN candidates. The fractions of PISNe at $z>6$ detected in multiple epochs in our survey simulations are 89\% (0.5~year cadence), 76\% (1.0~year cadence), and 44\% (1.5~year cadence). The fractions of SLSNe at $z>6$ detected in multiple epochs are 37\% (0.5~year cadence), 5\% (1.0~year cadence), and 0.1\% (1.5~year cadence). While we showed that we can identify high-redshift SNe in one epoch using the CDM with the F158 and F213 combination, the shorter cadence is preferred to acquire additional light-curve information for the efficient high-redshift SN screening.

\section{Summary}\label{sec:summary}
We investigated the best survey strategy to discover and identify PISNe and SLSNe at $z>6$ with \RST{}. We first determined the best filter combination to  distinguish efficiently high-redshift PISNe and SLSNe from Type~Ia and Type~II SNe at $z>1.5$ that may appear bright in a NIR color-magnitude diagram. 
We showed that the newly implemented \RST{} F213 filter is critical in identifying high-redshift PISNe and SLSNe. We found that F158 and F213 are the best combination to efficiently identify both high-redshift PISNe and SLSNe. Transient survey depths need to be $\mathrm{F158}=27.0~\mathrm{mag}$ and $\mathrm{F213}=26.5~\mathrm{mag}$ at each epoch for high-redshift SN identification. We also conducted mock survey simulations with a $10~\mathrm{deg^2}$ survey area with these filters and depths. We found that a 5-year survey with 1-year cadence can discover $22.5\pm2.8$ PISNe and $3.1\pm0.3$ SLSNe at $z>6$, depending on the uncertain cosmic SFR density. The same survey would discover $76.1\pm8.2$ PISNe and $9.1\pm0.9$ SLSNe at $5<z<6$.  Such a \RST{} survey requires approximately 525~hours in total observing time. The legacy data acquired in the survey would be beneficial for many different science cases, including the study of high-redshift galaxies.

Significant observed deviation from our computed estimates could imply that the high-redshift PISN/SLSN rates differ from what we assumed (Fig.~\ref{fig:snrate}), possibly because of a different cosmic SFR density and/or a different SN fraction. The causes of possible deviations can be informed by PISNe and SLSNe at $z < 6$ obtained by the same \RST{} survey or other transient surveys conducted by Euclid or optical facilities. If we simply extrapolate the high-redshift PISN rate we adopted to the nearby Universe, the nearby PISN rate would remain higher than the nearby SLSN rate. This is not likely and it is still unclear at what redshift PISNe start to appear efficiently. If we have no detections of PISNe with the proposed \RST{} transient survey, the PISN rate at $5 < z < 6$ would be lower than that we expect by a factor of around 100 and the PISN rate at $z > 6$ would be lower by a factor of around 10. Whatever the outcome, the proposed \RST{} survey would provide a significant constraint on the PISN event rate evolution at high redshifts.

Discovering SNe at $z>6$ requires high-sensitivity wide-field imagers in the NIR. \RST{} with the newly implemented K213 filter provides a very unique opportunity to search for the high-redshift SNe that have essential information on massive stars and their explosions at the reionization epoch. In addition, James Webb Space Telescope (JWST) may still be operated in the \RST{} era and could conduct spectroscopic observations of high-redshift PISN and SLSN candidates identified with the deep \RST{} transient survey we propose. The F158-F213 CDM allows us to identify efficiently high-redshift SN candidates and trigger the JWST follow-up observations. 
\RST{} provides an important capability for high-redshift SN discovery, and executing a time-domain survey like that presented
here in a \RST{} deep field would help maximize the science return of the facility.

\begin{acknowledgments}
We thank the JWST Cycle 1 high-redshift SN survey proposal team for helpful discussions.
TJM is supported by the Grants-in-Aid for Scientific Research of the Japan Society for the Promotion of Science (JP18K13585, JP20H00174, JP21K13966, JP21H04997).
BER acknowledges support from the NASA Roman Extragalactic Potential Observations Science Investigation Team contract NNG16PJ25C and grant 80NSSC18K0563.
This work was supported by JSPS Core-to-Core Program (JPJSCCA20210003).
\end{acknowledgments}

\bibliography{sample631}{}

\begin{thebibliography}{}
\expandafter\ifx\csname natexlab\endcsname\relax\def\natexlab#1{#1}\fi
\providecommand{\url}[1]{\href{#1}{#1}}
\providecommand{\dodoi}[1]{doi:~\href{http://doi.org/#1}{\nolinkurl{#1}}}
\providecommand{\doeprint}[1]{\href{http://ascl.net/#1}{\nolinkurl{http://ascl.net/#1}}}
\providecommand{\doarXiv}[1]{\href{https://arxiv.org/abs/#1}{\nolinkurl{https://arxiv.org/abs/#1}}}

\bibitem[{{Abel} {et~al.}(2002){Abel}, {Bryan}, \& {Norman}}]{abel2002}
{Abel}, T., {Bryan}, G.~L., \& {Norman}, M.~L. 2002, Science, 295, 93,
  \dodoi{10.1126/science.295.5552.93}

\bibitem[{{Angus} {et~al.}(2019){Angus}, {Smith}, {Sullivan}, {Inserra},
  {Wiseman}, {D'Andrea}, {Thomas}, {Nichol}, {Galbany}, {Childress}, {Asorey},
  {Brown}, {Casas}, {Castander}, {Curtin}, {Frohmaier}, {Glazebrook}, {Gruen},
  {Gutierrez}, {Kessler}, {Kim}, {Lidman}, {Macaulay}, {Nugent}, {Pursiainen},
  {Sako}, {Soares-Santos}, {Thomas}, {Abbott}, {Avila}, {Bertin}, {Brooks},
  {Buckley-Geer}, {Burke}, {Carnero Rosell}, {Carretero}, {da Costa}, {De
  Vicente}, {Desai}, {Diehl}, {Doel}, {Eifler}, {Flaugher}, {Fosalba},
  {Frieman}, {Garc{\'\i}a-Bellido}, {Gruendl}, {Gschwend}, {Hartley},
  {Hollowood}, {Honscheid}, {Hoyle}, {James}, {Kuehn}, {Kuropatkin}, {Lahav},
  {Lima}, {Maia}, {March}, {Marshall}, {Menanteau}, {Miller}, {Miquel},
  {Ogando}, {Plazas}, {Romer}, {Sanchez}, {Schindler}, {Schubnell}, {Sobreira},
  {Suchyta}, {Swanson}, {Tarle}, {Thomas}, {Tucker}, \& {DES
  Collaboration}}]{angus2019}
{Angus}, C.~R., {Smith}, M., {Sullivan}, M., {et~al.} 2019, \mnras, 487, 2215,
  \dodoi{10.1093/mnras/stz1321}

\bibitem[{{Aoki} {et~al.}(2014){Aoki}, {Tominaga}, {Beers}, {Honda}, \&
  {Lee}}]{aoki2014}
{Aoki}, W., {Tominaga}, N., {Beers}, T.~C., {Honda}, S., \& {Lee}, Y.~S. 2014,
  Science, 345, 912, \dodoi{10.1126/science.1252633}

\bibitem[{{Arnett}(1982)}]{arnett1982}
{Arnett}, W.~D. 1982, \apj, 253, 785, \dodoi{10.1086/159681}

\bibitem[{{Barkat} {et~al.}(1967){Barkat}, {Rakavy}, \&
  {Sack}}]{barkat1967pisn}
{Barkat}, Z., {Rakavy}, G., \& {Sack}, N. 1967, \prl, 18, 379,
  \dodoi{10.1103/PhysRevLett.18.379}

\bibitem[{{Blanchard} {et~al.}(2020){Blanchard}, {Berger}, {Nicholl}, \&
  {Villar}}]{blanchard2020}
{Blanchard}, P.~K., {Berger}, E., {Nicholl}, M., \& {Villar}, V.~A. 2020, \apj,
  897, 114, \dodoi{10.3847/1538-4357/ab9638}

\bibitem[{{Bose} {et~al.}(2018){Bose}, {Dong}, {Pastorello}, {Filippenko},
  {Kochanek}, {Mauerhan}, {Romero-Ca{\~n}izales}, {Brink}, {Chen}, {Prieto},
  {Post}, {Ashall}, {Grupe}, {Tomasella}, {Benetti}, {Shappee}, {Stanek},
  {Cai}, {Falco}, {Lundqvist}, {Mattila}, {Mutel}, {Ochner}, {Pooley},
  {Stritzinger}, {Villanueva}, {Zheng}, {Beswick}, {Brown}, {Cappellaro},
  {Davis}, {Fraser}, {de Jaeger}, {Elias-Rosa}, {Gall}, {Gaudi}, {Herczeg},
  {Hestenes}, {Holoien}, {Hosseinzadeh}, {Hsiao}, {Hu}, {Jaejin}, {Jeffers},
  {Koff}, {Kumar}, {Kurtenkov}, {Lau}, {Prentice}, {Reynolds}, {Rudy},
  {Shahbandeh}, {Somero}, {Stassun}, {Thompson}, {Valenti}, {Woo}, \&
  {Yunus}}]{bose2018}
{Bose}, S., {Dong}, S., {Pastorello}, A., {et~al.} 2018, \apj, 853, 57,
  \dodoi{10.3847/1538-4357/aaa298}

\bibitem[{{Bouwens} {et~al.}(2021){Bouwens}, {Oesch}, {Stefanon},
  {Illingworth}, {Labbe}, {Reddy}, {Atek}, {Montes}, {Naidu}, {Nanayakkara},
  {Nelson}, \& {Wilkins}}]{bouwens2021}
{Bouwens}, R.~J., {Oesch}, P.~A., {Stefanon}, M., {et~al.} 2021, arXiv
  e-prints, arXiv:2102.07775.
\newblock \doarXiv{2102.07775}

\bibitem[{{Bromm} \& {Yoshida}(2011)}]{bromm2011}
{Bromm}, V., \& {Yoshida}, N. 2011, \araa, 49, 373,
  \dodoi{10.1146/annurev-astro-081710-102608}

\bibitem[{{Chatzopoulos} {et~al.}(2015){Chatzopoulos}, {van Rossum}, {Craig},
  {Whalen}, {Smidt}, \& {Wiggins}}]{chatzopoulos2015pisnrot}
{Chatzopoulos}, E., {van Rossum}, D.~R., {Craig}, W.~J., {et~al.} 2015, \apj,
  799, 18, \dodoi{10.1088/0004-637X/799/1/18}

\bibitem[{{Chatzopoulos} {et~al.}(2012){Chatzopoulos}, {Wheeler}, \&
  {Vinko}}]{chatzopoulos2012}
{Chatzopoulos}, E., {Wheeler}, J.~C., \& {Vinko}, J. 2012, \apj, 746, 121,
  \dodoi{10.1088/0004-637X/746/2/121}

\bibitem[{{Chen} {et~al.}(2017){Chen}, {Smartt}, {Yates}, {Nicholl},
  {Kr{\"u}hler}, {Schady}, {Dennefeld}, \&
  {Inserra}}]{chen2017slsnhostmetallicity}
{Chen}, T.-W., {Smartt}, S.~J., {Yates}, R.~M., {et~al.} 2017, \mnras, 470,
  3566, \dodoi{10.1093/mnras/stx1428}

\bibitem[{{Chevalier} \& {Irwin}(2011)}]{chevalier2011}
{Chevalier}, R.~A., \& {Irwin}, C.~M. 2011, \apjl, 729, L6,
  \dodoi{10.1088/2041-8205/729/1/L6}

\bibitem[{{Cooke} {et~al.}(2012){Cooke}, {Sullivan}, {Gal-Yam}, {Barton},
  {Carlberg}, {Ryan-Weber}, {Horst}, {Omori}, \& {D{\'\i}az}}]{cooke2012}
{Cooke}, J., {Sullivan}, M., {Gal-Yam}, A., {et~al.} 2012, \nat, 491, 228,
  \dodoi{10.1038/nature11521}

\bibitem[{{Curtin} {et~al.}(2019){Curtin}, {Cooke}, {Moriya}, {Tanaka},
  {Quimby}, {Bernard}, {Galbany}, {Jiang}, {Lee}, {Maeda}, {Morokuma},
  {Nomoto}, {Pignata}, {Pritchard}, {Suzuki}, {Takahashi}, {Tanaka},
  {Tominaga}, {Yamaguchi}, \& {Yasuda}}]{curtin2019shizuca}
{Curtin}, C., {Cooke}, J., {Moriya}, T.~J., {et~al.} 2019, \apjs, 241, 17,
  \dodoi{10.3847/1538-4365/ab07c8}

\bibitem[{{de Bennassuti} {et~al.}(2017){de Bennassuti}, {Salvadori},
  {Schneider}, {Valiante}, \& {Omukai}}]{debennassuti2017}
{de Bennassuti}, M., {Salvadori}, S., {Schneider}, R., {Valiante}, R., \&
  {Omukai}, K. 2017, \mnras, 465, 926, \dodoi{10.1093/mnras/stw2687}

\bibitem[{{De Cia} {et~al.}(2018){De Cia}, {Gal-Yam}, {Rubin}, {Leloudas},
  {Vreeswijk}, {Perley}, {Quimby}, {Yan}, {Sullivan}, {Fl{\"o}rs}, {Sollerman},
  {Bersier}, {Cenko}, {Gal-Yam}, {Maguire}, {Ofek}, {Prentice}, {Schulze},
  {Spyromilio}, {Valenti}, {Arcavi}, {Corsi}, {Howell}, {Mazzali}, {Kasliwal},
  {Taddia}, \& {Yaron}}]{decia2018}
{De Cia}, A., {Gal-Yam}, A., {Rubin}, A., {et~al.} 2018, \apj, 860, 100,
  \dodoi{10.3847/1538-4357/aab9b6}

\bibitem[{{de Souza} {et~al.}(2013){de Souza}, {Ishida}, {Johnson}, {Whalen},
  \& {Mesinger}}]{desouza2013}
{de Souza}, R.~S., {Ishida}, E.~E.~O., {Johnson}, J.~L., {Whalen}, D.~J., \&
  {Mesinger}, A. 2013, \mnras, 436, 1555, \dodoi{10.1093/mnras/stt1680}

\bibitem[{{Dessart} {et~al.}(2012){Dessart}, {Hillier}, {Waldman}, {Livne}, \&
  {Blondin}}]{dessart2012}
{Dessart}, L., {Hillier}, D.~J., {Waldman}, R., {Livne}, E., \& {Blondin}, S.
  2012, \mnras, 426, L76, \dodoi{10.1111/j.1745-3933.2012.01329.x}

\bibitem[{{Dessart} {et~al.}(2013){Dessart}, {Waldman}, {Livne}, {Hillier}, \&
  {Blondin}}]{dessart2013pisn}
{Dessart}, L., {Waldman}, R., {Livne}, E., {Hillier}, D.~J., \& {Blondin}, S.
  2013, \mnras, 428, 3227, \dodoi{10.1093/mnras/sts269}

\bibitem[{{du Buisson} {et~al.}(2020){du Buisson}, {Marchant}, {Podsiadlowski},
  {Kobayashi}, {Abdalla}, {Taylor}, {Mandel}, {de Mink}, {Moriya}, \&
  {Langer}}]{dubuisson2020}
{du Buisson}, L., {Marchant}, P., {Podsiadlowski}, P., {et~al.} 2020, \mnras,
  499, 5941, \dodoi{10.1093/mnras/staa3225}

\bibitem[{{Dunlop} {et~al.}(2012){Dunlop}, {McLure}, {Robertson}, {Ellis},
  {Stark}, {Cirasuolo}, \& {de Ravel}}]{dunlop2012}
{Dunlop}, J.~S., {McLure}, R.~J., {Robertson}, B.~E., {et~al.} 2012, \mnras,
  420, 901, \dodoi{10.1111/j.1365-2966.2011.20102.x}

\bibitem[{{Dunlop} {et~al.}(2013){Dunlop}, {Rogers}, {McLure}, {Ellis},
  {Robertson}, {Koekemoer}, {Dayal}, {Curtis-Lake}, {Wild}, {Charlot},
  {Bowler}, {Schenker}, {Ouchi}, {Ono}, {Cirasuolo}, {Furlanetto}, {Stark},
  {Targett}, \& {Schneider}}]{dunlop2013}
{Dunlop}, J.~S., {Rogers}, A.~B., {McLure}, R.~J., {et~al.} 2013, \mnras, 432,
  3520, \dodoi{10.1093/mnras/stt702}

\bibitem[{{Eldridge} {et~al.}(2013){Eldridge}, {Fraser}, {Smartt}, {Maund}, \&
  {Crockett}}]{eldridge2013}
{Eldridge}, J.~J., {Fraser}, M., {Smartt}, S.~J., {Maund}, J.~R., \&
  {Crockett}, R.~M. 2013, \mnras, 436, 774, \dodoi{10.1093/mnras/stt1612}

\bibitem[{{Frohmaier} {et~al.}(2021){Frohmaier}, {Angus}, {Vincenzi},
  {Sullivan}, {Smith}, {Nugent}, {Cenko}, {Gal-Yam}, {Kulkarni}, {Law}, \&
  {Quimby}}]{frohmaier2021}
{Frohmaier}, C., {Angus}, C.~R., {Vincenzi}, M., {et~al.} 2021, \mnras, 500,
  5142, \dodoi{10.1093/mnras/staa3607}

\bibitem[{{Gal-Yam}(2019)}]{gal-yam2019}
{Gal-Yam}, A. 2019, \araa, 57, 305, \dodoi{10.1146/annurev-astro-081817-051819}

\bibitem[{{Gal-Yam} {et~al.}(2009){Gal-Yam}, {Mazzali}, {Ofek}, {Nugent},
  {Kulkarni}, {Kasliwal}, {Quimby}, {Filippenko}, {Cenko}, {Chornock},
  {Waldman}, {Kasen}, {Sullivan}, {Beshore}, {Drake}, {Thomas}, {Bloom},
  {Poznanski}, {Miller}, {Foley}, {Silverman}, {Arcavi}, {Ellis}, \&
  {Deng}}]{gal-yam2009sn2007bi}
{Gal-Yam}, A., {Mazzali}, P., {Ofek}, E.~O., {et~al.} 2009, \nat, 462, 624,
  \dodoi{10.1038/nature08579}

\bibitem[{{Georgy} {et~al.}(2017){Georgy}, {Meynet}, {Ekstr{\"o}m}, {Wade},
  {Petit}, {Keszthelyi}, \& {Hirschi}}]{georgy2017}
{Georgy}, C., {Meynet}, G., {Ekstr{\"o}m}, S., {et~al.} 2017, \aap, 599, L5,
  \dodoi{10.1051/0004-6361/201730401}

\bibitem[{{Gilmer} {et~al.}(2017){Gilmer}, {Kozyreva}, {Hirschi},
  {Fr{\"o}hlich}, \& {Yusof}}]{gilmer2017pisn}
{Gilmer}, M.~S., {Kozyreva}, A., {Hirschi}, R., {Fr{\"o}hlich}, C., \& {Yusof},
  N. 2017, \apj, 846, 100, \dodoi{10.3847/1538-4357/aa8461}

\bibitem[{{Goldstein} {et~al.}(2016){Goldstein}, {Connaughton}, {Briggs}, \&
  {Burns}}]{goldstein2016}
{Goldstein}, A., {Connaughton}, V., {Briggs}, M.~S., \& {Burns}, E. 2016, \apj,
  818, 18, \dodoi{10.3847/0004-637X/818/1/18}

\bibitem[{{G{\"o}tberg} {et~al.}(2017){G{\"o}tberg}, {de Mink}, \&
  {Groh}}]{gotberg2017}
{G{\"o}tberg}, Y., {de Mink}, S.~E., \& {Groh}, J.~H. 2017, \aap, 608, A11,
  \dodoi{10.1051/0004-6361/201730472}

\bibitem[{{G{\"o}tberg} {et~al.}(2020){G{\"o}tberg}, {de Mink}, {McQuinn},
  {Zapartas}, {Groh}, \& {Norman}}]{gotberg2020}
{G{\"o}tberg}, Y., {de Mink}, S.~E., {McQuinn}, M., {et~al.} 2020, \aap, 634,
  A134, \dodoi{10.1051/0004-6361/201936669}

\bibitem[{{Hartwig} {et~al.}(2018){Hartwig}, {Bromm}, \& {Loeb}}]{hartwig2018}
{Hartwig}, T., {Bromm}, V., \& {Loeb}, A. 2018, \mnras, 479, 2202,
  \dodoi{10.1093/mnras/sty1576}

\bibitem[{{Heger} \& {Woosley}(2002)}]{heger2002}
{Heger}, A., \& {Woosley}, S.~E. 2002, \apj, 567, 532, \dodoi{10.1086/338487}

\bibitem[{{Higgins} {et~al.}(2021){Higgins}, {Sander}, {Vink}, \&
  {Hirschi}}]{higgins2021}
{Higgins}, E.~R., {Sander}, A. A.~C., {Vink}, J.~S., \& {Hirschi}, R. 2021,
  arXiv e-prints, arXiv:2105.12139.
\newblock \doarXiv{2105.12139}

\bibitem[{{Hirano} {et~al.}(2015){Hirano}, {Hosokawa}, {Yoshida}, {Omukai}, \&
  {Yorke}}]{hirano2015}
{Hirano}, S., {Hosokawa}, T., {Yoshida}, N., {Omukai}, K., \& {Yorke}, H.~W.
  2015, \mnras, 448, 568, \dodoi{10.1093/mnras/stv044}

\bibitem[{{Hounsell} {et~al.}(2018){Hounsell}, {Scolnic}, {Foley}, {Kessler},
  {Miranda}, {Avelino}, {Bohlin}, {Filippenko}, {Frieman}, {Jha}, {Kelly},
  {Kirshner}, {Mandel}, {Rest}, {Riess}, {Rodney}, \&
  {Strolger}}]{hounsell2018}
{Hounsell}, R., {Scolnic}, D., {Foley}, R.~J., {et~al.} 2018, \apj, 867, 23,
  \dodoi{10.3847/1538-4357/aac08b}

\bibitem[{{Hsiao} {et~al.}(2007){Hsiao}, {Conley}, {Howell}, {Sullivan},
  {Pritchet}, {Carlberg}, {Nugent}, \& {Phillips}}]{hsiao2007}
{Hsiao}, E.~Y., {Conley}, A., {Howell}, D.~A., {et~al.} 2007, \apj, 663, 1187,
  \dodoi{10.1086/518232}

\bibitem[{{Hsu} {et~al.}(2021){Hsu}, {Hosseinzadeh}, \& {Berger}}]{hsu2021}
{Hsu}, B., {Hosseinzadeh}, G., \& {Berger}, E. 2021, arXiv e-prints,
  arXiv:2104.09639.
\newblock \doarXiv{2104.09639}

\bibitem[{{Inserra} \& {Smartt}(2014)}]{inserra2014}
{Inserra}, C., \& {Smartt}, S.~J. 2014, \apj, 796, 87,
  \dodoi{10.1088/0004-637X/796/2/87}

\bibitem[{{Inserra} {et~al.}(2013){Inserra}, {Smartt}, {Jerkstrand}, {Valenti},
  {Fraser}, {Wright}, {Smith}, {Chen}, {Kotak}, {Pastorello}, {Nicholl},
  {Bresolin}, {Kudritzki}, {Benetti}, {Botticella}, {Burgett}, {Chambers},
  {Ergon}, {Flewelling}, {Fynbo}, {Geier}, {Hodapp}, {Howell}, {Huber},
  {Kaiser}, {Leloudas}, {Magill}, {Magnier}, {McCrum}, {Metcalfe}, {Price},
  {Rest}, {Sollerman}, {Sweeney}, {Taddia}, {Taubenberger}, {Tonry},
  {Wainscoat}, {Waters}, \& {Young}}]{inserra2013}
{Inserra}, C., {Smartt}, S.~J., {Jerkstrand}, A., {et~al.} 2013, \apj, 770,
  128, \dodoi{10.1088/0004-637X/770/2/128}

\bibitem[{{Inserra} {et~al.}(2018){Inserra}, {Nichol}, {Scovacricchi},
  {Amiaux}, {Brescia}, {Burigana}, {Cappellaro}, {Carvalho}, {Cavuoti},
  {Conforti}, {Cuillandre}, {da Silva}, {De Rosa}, {Della Valle}, {Dinis},
  {Franceschi}, {Hook}, {Hudelot}, {Jahnke}, {Kitching}, {Kurki-Suonio},
  {Lloro}, {Longo}, {Maiorano}, {Maris}, {Rhodes}, {Scaramella}, {Smartt},
  {Sullivan}, {Tao}, {Toledo-Moreo}, {Tereno}, {Trifoglio}, \&
  {Valenziano}}]{inserra2018}
{Inserra}, C., {Nichol}, R.~C., {Scovacricchi}, D., {et~al.} 2018, \aap, 609,
  A83, \dodoi{10.1051/0004-6361/201731758}

\bibitem[{{Inserra} {et~al.}(2021){Inserra}, {Sullivan}, {Angus}, {Macaulay},
  {Nichol}, {Smith}, {Frohmaier}, {Guti{\'e}rrez}, {Vicenzi}, {M{\"o}ller},
  {Brout}, {Brown}, {Davis}, {D'Andrea}, {Galbany}, {Kessler}, {Kim}, {Pan},
  {Pursiainen}, {Scolnic}, {Thomas}, {Wiseman}, {Abbott}, {Annis}, {Avila},
  {Bertin}, {Brooks}, {Burke}, {Carnero Rosell}, {Carrasco Kind}, {Carretero},
  {Castander}, {Cawthon}, {Desai}, {Diehl}, {Eifler}, {Finley}, {Flaugher},
  {Fosalba}, {Frieman}, {Garcia-Bellido}, {Gaztanaga}, {Gerdes},
  {Giannantonio}, {Gruen}, {Gruendl}, {Gschwend}, {Gutierrez}, {Hollowood},
  {Honscheid}, {James}, {Krause}, {Kuehn}, {Kuropatkin}, {Li}, {Lidman},
  {Lima}, {Maia}, {Marshall}, {Martini}, {Menanteau}, {Miquel}, {Plazas
  Malag{\'o}n}, {Romer}, {Roodman}, {Sako}, {Sanchez}, {Scarpine}, {Schubnell},
  {Serrano}, {Sevilla-Noarbe}, {Soares-Santos}, {Sobreira}, {Suchyta},
  {Swanson}, {Tarle}, {Thomas}, {Tucker}, {Vikram}, {Walker}, {Zhang},
  {Asorey}, {Calcino}, {Carollo}, {Glazebrook}, {Hinton}, {Hoormann}, {Lewis},
  {Sharp}, {Swann}, {Tucker}, \& {DES Collaboration}}]{inserra2021}
{Inserra}, C., {Sullivan}, M., {Angus}, C.~R., {et~al.} 2021, \mnras, 504,
  2535, \dodoi{10.1093/mnras/stab978}

\bibitem[{{Jerkstrand} {et~al.}(2020){Jerkstrand}, {Maeda}, \&
  {Kawabata}}]{jerkstrand2020}
{Jerkstrand}, A., {Maeda}, K., \& {Kawabata}, K.~S. 2020, Science, 367, 415,
  \dodoi{10.1126/science.aaw1469}

\bibitem[{{Jerkstrand} {et~al.}(2016){Jerkstrand}, {Smartt}, \&
  {Heger}}]{jerkstrand2016pisnneb}
{Jerkstrand}, A., {Smartt}, S.~J., \& {Heger}, A. 2016, \mnras, 455, 3207,
  \dodoi{10.1093/mnras/stv2369}

\bibitem[{{Kasen} \& {Bildsten}(2010)}]{kasen2010}
{Kasen}, D., \& {Bildsten}, L. 2010, \apj, 717, 245,
  \dodoi{10.1088/0004-637X/717/1/245}

\bibitem[{{Kasen} {et~al.}(2011){Kasen}, {Woosley}, \& {Heger}}]{kasen2011}
{Kasen}, D., {Woosley}, S.~E., \& {Heger}, A. 2011, \apj, 734, 102,
  \dodoi{10.1088/0004-637X/734/2/102}

\bibitem[{{Kozyreva} {et~al.}(2014){Kozyreva}, {Blinnikov}, {Langer}, \&
  {Yoon}}]{kozyreva2014pisnprop}
{Kozyreva}, A., {Blinnikov}, S., {Langer}, N., \& {Yoon}, S.~C. 2014, \aap,
  565, A70, \dodoi{10.1051/0004-6361/201423447}

\bibitem[{{Langer} {et~al.}(2007){Langer}, {Norman}, {de Koter}, {Vink},
  {Cantiello}, \& {Yoon}}]{langer2007}
{Langer}, N., {Norman}, C.~A., {de Koter}, A., {et~al.} 2007, \aap, 475, L19,
  \dodoi{10.1051/0004-6361:20078482}

\bibitem[{{Laplace} {et~al.}(2021){Laplace}, {Justham}, {Renzo}, {G{\"o}tberg},
  {Farmer}, {Vartanyan}, \& {de Mink}}]{laplace2021}
{Laplace}, E., {Justham}, S., {Renzo}, M., {et~al.} 2021, arXiv e-prints,
  arXiv:2102.05036.
\newblock \doarXiv{2102.05036}

\bibitem[{{Leloudas} {et~al.}(2015){Leloudas}, {Schulze}, {Kr{\"u}hler},
  {Gorosabel}, {Christensen}, {Mehner}, {de Ugarte Postigo}, {Amor{\'\i}n},
  {Th{\"o}ne}, {Anderson}, {Bauer}, {Gallazzi}, {He{\l}miniak}, {Hjorth},
  {Ibar}, {Malesani}, {Morell}, {Vinko}, \& {Wheeler}}]{leloudas2015}
{Leloudas}, G., {Schulze}, S., {Kr{\"u}hler}, T., {et~al.} 2015, \mnras, 449,
  917, \dodoi{10.1093/mnras/stv320}

\bibitem[{{Liu} {et~al.}(2017){Liu}, {Wang}, {Wang}, {Dai}, {Yu}, \&
  {Peng}}]{liu2017}
{Liu}, L.-D., {Wang}, S.-Q., {Wang}, L.-J., {et~al.} 2017, \apj, 842, 26,
  \dodoi{10.3847/1538-4357/aa73d9}

\bibitem[{{Lunnan} {et~al.}(2018){Lunnan}, {Chornock}, {Berger}, {Jones},
  {Rest}, {Czekala}, {Dittmann}, {Drout}, {Foley}, {Fong}, {Kirshner},
  {Laskar}, {Leibler}, {Margutti}, {Milisavljevic}, {Narayan}, {Pan}, {Riess},
  {Roth}, {Sanders}, {Scolnic}, {Smartt}, {Smith}, {Chambers}, {Draper},
  {Flewelling}, {Huber}, {Kaiser}, {Kudritzki}, {Magnier}, {Metcalfe},
  {Wainscoat}, {Waters}, \& {Willman}}]{lunnan2018pansample}
{Lunnan}, R., {Chornock}, R., {Berger}, E., {et~al.} 2018, \apj, 852, 81,
  \dodoi{10.3847/1538-4357/aa9f1a}

\bibitem[{{Machida} {et~al.}(2009){Machida}, {Omukai}, {Matsumoto}, \&
  {Inutsuka}}]{machida2009}
{Machida}, M.~N., {Omukai}, K., {Matsumoto}, T., \& {Inutsuka}, S.-I. 2009,
  \mnras, 399, 1255, \dodoi{10.1111/j.1365-2966.2009.15394.x}

\bibitem[{{Madau} \& {Dickinson}(2014)}]{madau2014}
{Madau}, P., \& {Dickinson}, M. 2014, \araa, 52, 415,
  \dodoi{10.1146/annurev-astro-081811-125615}

\bibitem[{{Marchant} {et~al.}(2016){Marchant}, {Langer}, {Podsiadlowski},
  {Tauris}, \& {Moriya}}]{marchant2016}
{Marchant}, P., {Langer}, N., {Podsiadlowski}, P., {Tauris}, T.~M., \&
  {Moriya}, T.~J. 2016, \aap, 588, A50, \dodoi{10.1051/0004-6361/201628133}

\bibitem[{{Mazzali} {et~al.}(2019){Mazzali}, {Moriya}, {Tanaka}, \&
  {Woosley}}]{mazzali2019}
{Mazzali}, P.~A., {Moriya}, T.~J., {Tanaka}, M., \& {Woosley}, S.~E. 2019,
  \mnras, 484, 3451, \dodoi{10.1093/mnras/stz177}

\bibitem[{{Mokiem} {et~al.}(2007){Mokiem}, {de Koter}, {Vink}, {Puls}, {Evans},
  {Smartt}, {Crowther}, {Herrero}, {Langer}, {Lennon}, {Najarro}, \&
  {Villamariz}}]{mokiem2007}
{Mokiem}, M.~R., {de Koter}, A., {Vink}, J.~S., {et~al.} 2007, \aap, 473, 603,
  \dodoi{10.1051/0004-6361:20077545}

\bibitem[{{Moriya} {et~al.}(2013){Moriya}, {Blinnikov}, {Tominaga}, {Yoshida},
  {Tanaka}, {Maeda}, \& {Nomoto}}]{moriya2013sn2006gy}
{Moriya}, T.~J., {Blinnikov}, S.~I., {Tominaga}, N., {et~al.} 2013, \mnras,
  428, 1020, \dodoi{10.1093/mnras/sts075}

\bibitem[{{Moriya} \& {Langer}(2015)}]{moriya2015pisn}
{Moriya}, T.~J., \& {Langer}, N. 2015, \aap, 573, A18,
  \dodoi{10.1051/0004-6361/201424957}

\bibitem[{{Moriya} {et~al.}(2019{\natexlab{a}}){Moriya}, {Mazzali}, \&
  {Tanaka}}]{moriya2019sn2007bi}
{Moriya}, T.~J., {Mazzali}, P.~A., \& {Tanaka}, M. 2019{\natexlab{a}}, \mnras,
  484, 3443, \dodoi{10.1093/mnras/stz262}

\bibitem[{{Moriya} {et~al.}(2018){Moriya}, {Sorokina}, \&
  {Chevalier}}]{moriya2018slsnreview}
{Moriya}, T.~J., {Sorokina}, E.~I., \& {Chevalier}, R.~A. 2018, \ssr, 214, 59,
  \dodoi{10.1007/s11214-018-0493-6}

\bibitem[{{Moriya} {et~al.}(2019{\natexlab{b}}){Moriya}, {Wong}, {Koyama},
  {Tanaka}, {Oguri}, {Hilbert}, \& {Nomoto}}]{moriya2019ultimate}
{Moriya}, T.~J., {Wong}, K.~C., {Koyama}, Y., {et~al.} 2019{\natexlab{b}},
  \pasj, 71, 59, \dodoi{10.1093/pasj/psz035}

\bibitem[{{Moriya} {et~al.}(2019{\natexlab{c}}){Moriya}, {Tanaka}, {Yasuda},
  {Jiang}, {Lee}, {Maeda}, {Morokuma}, {Nomoto}, {Quimby}, {Suzuki},
  {Takahashi}, {Tanaka}, {Tominaga}, {Yamaguchi}, {Bernard}, {Cooke}, {Curtin},
  {Galbany}, {Gonz{\'a}lez-Gait{\'a}n}, {Pignata}, {Pritchard}, {Komiyama}, \&
  {Lupton}}]{moriya2019shizuca}
{Moriya}, T.~J., {Tanaka}, M., {Yasuda}, N., {et~al.} 2019{\natexlab{c}},
  \apjs, 241, 16, \dodoi{10.3847/1538-4365/ab07c5}

\bibitem[{{Moriya} {et~al.}(2021){Moriya}, {Jiang}, {Yasuda}, {Kokubo},
  {Kawana}, {Maeda}, {Pan}, {Quimby}, {Suzuki}, {Takahashi}, {Tanaka},
  {Tominaga}, {Nomoto}, {Cooke}, {Galbany}, {Gonz{\'a}lez-Gait{\'a}n}, {Lee},
  \& {Pignata}}]{moriya2021hsc}
{Moriya}, T.~J., {Jiang}, J.-a., {Yasuda}, N., {et~al.} 2021, \apj, 908, 249,
  \dodoi{10.3847/1538-4357/abcfc0}

\bibitem[{{Neill} {et~al.}(2011){Neill}, {Sullivan}, {Gal-Yam}, {Quimby},
  {Ofek}, {Wyder}, {Howell}, {Nugent}, {Seibert}, {Martin}, {Overzier},
  {Barlow}, {Foster}, {Friedman}, {Morrissey}, {Neff}, {Schiminovich},
  {Bianchi}, {Donas}, {Heckman}, {Lee}, {Madore}, {Milliard}, {Rich}, \&
  {Szalay}}]{neill2011}
{Neill}, J.~D., {Sullivan}, M., {Gal-Yam}, A., {et~al.} 2011, \apj, 727, 15,
  \dodoi{10.1088/0004-637X/727/1/15}

\bibitem[{{Nicholl} {et~al.}(2019){Nicholl}, {Berger}, {Blanchard}, {Gomez}, \&
  {Chornock}}]{nicholl2019}
{Nicholl}, M., {Berger}, E., {Blanchard}, P.~K., {Gomez}, S., \& {Chornock}, R.
  2019, \apj, 871, 102, \dodoi{10.3847/1538-4357/aaf470}

\bibitem[{{Nicholl} {et~al.}(2017{\natexlab{a}}){Nicholl}, {Berger},
  {Margutti}, {Blanchard}, {Guillochon}, {Leja}, \&
  {Chornock}}]{nicholl2017sn2017egm}
{Nicholl}, M., {Berger}, E., {Margutti}, R., {et~al.} 2017{\natexlab{a}},
  \apjl, 845, L8, \dodoi{10.3847/2041-8213/aa82b1}

\bibitem[{{Nicholl} {et~al.}(2017{\natexlab{b}}){Nicholl}, {Berger},
  {Margutti}, {Blanchard}, {Milisavljevic}, {Challis}, {Metzger}, \&
  {Chornock}}]{nicholl2017gaia16apd}
---. 2017{\natexlab{b}}, \apjl, 835, L8, \dodoi{10.3847/2041-8213/aa56c5}

\bibitem[{{Nicholl} {et~al.}(2017{\natexlab{c}}){Nicholl}, {Guillochon}, \&
  {Berger}}]{nicholl2017}
{Nicholl}, M., {Guillochon}, J., \& {Berger}, E. 2017{\natexlab{c}}, \apj, 850,
  55, \dodoi{10.3847/1538-4357/aa9334}

\bibitem[{{Nicholl} {et~al.}(2013){Nicholl}, {Smartt}, {Jerkstrand}, {Inserra},
  {McCrum}, {Kotak}, {Fraser}, {Wright}, {Chen}, {Smith}, {Young}, {Sim},
  {Valenti}, {Howell}, {Bresolin}, {Kudritzki}, {Tonry}, {Huber}, {Rest},
  {Pastorello}, {Tomasella}, {Cappellaro}, {Benetti}, {Mattila}, {Kankare},
  {Kangas}, {Leloudas}, {Sollerman}, {Taddia}, {Berger}, {Chornock}, {Narayan},
  {Stubbs}, {Foley}, {Lunnan}, {Soderberg}, {Sanders}, {Milisavljevic},
  {Margutti}, {Kirshner}, {Elias-Rosa}, {Morales-Garoffolo}, {Taubenberger},
  {Botticella}, {Gezari}, {Urata}, {Rodney}, {Riess}, {Scolnic}, {Wood-Vasey},
  {Burgett}, {Chambers}, {Flewelling}, {Magnier}, {Kaiser}, {Metcalfe},
  {Morgan}, {Price}, {Sweeney}, \& {Waters}}]{nicholl2013}
{Nicholl}, M., {Smartt}, S.~J., {Jerkstrand}, A., {et~al.} 2013, \nat, 502,
  346, \dodoi{10.1038/nature12569}

\bibitem[{{Nicholl} {et~al.}(2015){Nicholl}, {Smartt}, {Jerkstrand}, {Inserra},
  {Sim}, {Chen}, {Benetti}, {Fraser}, {Gal-Yam}, {Kankare}, {Maguire}, {Smith},
  {Sullivan}, {Valenti}, {Young}, {Baltay}, {Bauer}, {Baumont}, {Bersier},
  {Botticella}, {Childress}, {Dennefeld}, {Della Valle}, {Elias-Rosa},
  {Feindt}, {Galbany}, {Hadjiyska}, {Le Guillou}, {Leloudas}, {Mazzali},
  {McKinnon}, {Polshaw}, {Rabinowitz}, {Rostami}, {Scalzo}, {Schmidt},
  {Schulze}, {Sollerman}, {Taddia}, \& {Yuan}}]{nicholl2015}
---. 2015, \mnras, 452, 3869, \dodoi{10.1093/mnras/stv1522}

\bibitem[{{Nomoto} {et~al.}(2013){Nomoto}, {Kobayashi}, \&
  {Tominaga}}]{nomoto2013}
{Nomoto}, K., {Kobayashi}, C., \& {Tominaga}, N. 2013, \araa, 51, 457,
  \dodoi{10.1146/annurev-astro-082812-140956}

\bibitem[{{Nomoto} {et~al.}(1995){Nomoto}, {Iwamoto}, \& {Suzuki}}]{nomoto1995}
{Nomoto}, K.~I., {Iwamoto}, K., \& {Suzuki}, T. 1995, \physrep, 256, 173,
  \dodoi{10.1016/0370-1573(94)00107-E}

\bibitem[{{Pan} {et~al.}(2012){Pan}, {Kasen}, \& {Loeb}}]{pan2012}
{Pan}, T., {Kasen}, D., \& {Loeb}, A. 2012, \mnras, 422, 2701,
  \dodoi{10.1111/j.1365-2966.2012.20837.x}

\bibitem[{{Perley} {et~al.}(2016){Perley}, {Quimby}, {Yan}, {Vreeswijk}, {De
  Cia}, {Lunnan}, {Gal-Yam}, {Yaron}, {Filippenko}, {Graham}, {Laher}, \&
  {Nugent}}]{perley2016}
{Perley}, D.~A., {Quimby}, R.~M., {Yan}, L., {et~al.} 2016, \apj, 830, 13,
  \dodoi{10.3847/0004-637X/830/1/13}

\bibitem[{{Podsiadlowski} {et~al.}(1992){Podsiadlowski}, {Joss}, \&
  {Hsu}}]{podsiadlowski1992}
{Podsiadlowski}, P., {Joss}, P.~C., \& {Hsu}, J.~J.~L. 1992, \apj, 391, 246,
  \dodoi{10.1086/171341}

\bibitem[{{Prajs} {et~al.}(2017){Prajs}, {Sullivan}, {Smith}, {Levan},
  {Karpenka}, {Edwards}, {Walker}, {Wolf}, {Balland}, {Carlberg}, {Howell},
  {Lidman}, {Pain}, {Pritchet}, \& {Ruhlmann-Kleider}}]{prajs2017}
{Prajs}, S., {Sullivan}, M., {Smith}, M., {et~al.} 2017, \mnras, 464, 3568,
  \dodoi{10.1093/mnras/stw1942}

\bibitem[{{Quimby} {et~al.}(2013){Quimby}, {Yuan}, {Akerlof}, \&
  {Wheeler}}]{quimby2013}
{Quimby}, R.~M., {Yuan}, F., {Akerlof}, C., \& {Wheeler}, J.~C. 2013, \mnras,
  431, 912, \dodoi{10.1093/mnras/stt213}

\bibitem[{{Quimby} {et~al.}(2011){Quimby}, {Kulkarni}, {Kasliwal}, {Gal-Yam},
  {Arcavi}, {Sullivan}, {Nugent}, {Thomas}, {Howell}, {Nakar}, {Bildsten},
  {Theissen}, {Law}, {Dekany}, {Rahmer}, {Hale}, {Smith}, {Ofek}, {Zolkower},
  {Velur}, {Walters}, {Henning}, {Bui}, {McKenna}, {Poznanski}, {Cenko}, \&
  {Levitan}}]{quimby2011}
{Quimby}, R.~M., {Kulkarni}, S.~R., {Kasliwal}, M.~M., {et~al.} 2011, \nat,
  474, 487, \dodoi{10.1038/nature10095}

\bibitem[{{Rakavy} \& {Shaviv}(1967)}]{rakavy1967pisn}
{Rakavy}, G., \& {Shaviv}, G. 1967, \apj, 148, 803, \dodoi{10.1086/149204}

\bibitem[{{Reg{\H{o}}s} {et~al.}(2020){Reg{\H{o}}s}, {Vink{\'o}}, \&
  {Ziegler}}]{regos2020}
{Reg{\H{o}}s}, E., {Vink{\'o}}, J., \& {Ziegler}, B.~L. 2020, \apj, 894, 94,
  \dodoi{10.3847/1538-4357/ab8636}

\bibitem[{{Richardson} {et~al.}(2014){Richardson}, {Jenkins}, {Wright}, \&
  {Maddox}}]{richardson2014}
{Richardson}, D., {Jenkins}, Robert~L., I., {Wright}, J., \& {Maddox}, L. 2014,
  \aj, 147, 118, \dodoi{10.1088/0004-6256/147/5/118}

\bibitem[{{Robertson} \& {Ellis}(2012)}]{robertson2012}
{Robertson}, B.~E., \& {Ellis}, R.~S. 2012, \apj, 744, 95,
  \dodoi{10.1088/0004-637X/744/2/95}

\bibitem[{{Robertson} {et~al.}(2015){Robertson}, {Ellis}, {Furlanetto}, \&
  {Dunlop}}]{robertson2015}
{Robertson}, B.~E., {Ellis}, R.~S., {Furlanetto}, S.~R., \& {Dunlop}, J.~S.
  2015, \apjl, 802, L19, \dodoi{10.1088/2041-8205/802/2/L19}

\bibitem[{{Sander} \& {Vink}(2020)}]{sander2020b}
{Sander}, A. A.~C., \& {Vink}, J.~S. 2020, \mnras, 499, 873,
  \dodoi{10.1093/mnras/staa2712}

\bibitem[{{Sander} {et~al.}(2020){Sander}, {Vink}, \& {Hamann}}]{sander2020a}
{Sander}, A. A.~C., {Vink}, J.~S., \& {Hamann}, W.~R. 2020, \mnras, 491, 4406,
  \dodoi{10.1093/mnras/stz3064}

\bibitem[{{Scannapieco} {et~al.}(2005){Scannapieco}, {Madau}, {Woosley},
  {Heger}, \& {Ferrara}}]{scannapieco2005}
{Scannapieco}, E., {Madau}, P., {Woosley}, S., {Heger}, A., \& {Ferrara}, A.
  2005, \apj, 633, 1031, \dodoi{10.1086/444450}

\bibitem[{{Schulze} {et~al.}(2018){Schulze}, {Kr{\"u}hler}, {Leloudas},
  {Gorosabel}, {Mehner}, {Buchner}, {Kim}, {Ibar}, {Amor{\'\i}n},
  {Herrero-Illana}, {Anderson}, {Bauer}, {Christensen}, {de Pasquale}, {de
  Ugarte Postigo}, {Gallazzi}, {Hjorth}, {Morrell}, {Malesani}, {Sparre},
  {Stalder}, {Stark}, {Th{\"o}ne}, \& {Wheeler}}]{schulze2018}
{Schulze}, S., {Kr{\"u}hler}, T., {Leloudas}, G., {et~al.} 2018, \mnras, 473,
  1258, \dodoi{10.1093/mnras/stx2352}

\bibitem[{{Schulze} {et~al.}(2020){Schulze}, {Yaron}, {Sollerman}, {Leloudas},
  {Gal}, {Wright}, {Lunnan}, {Gal-Yam}, {Ofek}, {Perley}, {Filippenko},
  {Kasliwal}, {Kulkarni}, {Nugent}, {Quimby}, {Sullivan}, {Linn Strothjohann},
  {Arcavi}, {Ben-Ami}, {Bianco}, {Bloom}, {De}, {Fraser}, {Fremling}, {Horesh},
  {Johansson}, {Kelly}, {Knezevic}, {Maguire}, {Nyholm}, {Semeli
  Papadogiannakis}, {Petrushevska}, {Rubin}, {Yan}, {Yang}, {Adams}, {Bufano},
  {Clubb}, {Foley}, {Green}, {Harmanen}, {Ho}, {Hook}, {Hosseinzadeh},
  {Howell}, {Kong}, {Kotak}, {Matheson}, {McCully}, {Milisavljevic}, {Pan},
  {Poznanski}, {Shivvers}, \& {van Velzen}}]{schulze2020}
{Schulze}, S., {Yaron}, O., {Sollerman}, J., {et~al.} 2020, arXiv e-prints,
  arXiv:2008.05988.
\newblock \doarXiv{2008.05988}

\bibitem[{{Smith}(2014)}]{smith2014}
{Smith}, N. 2014, \araa, 52, 487, \dodoi{10.1146/annurev-astro-081913-040025}

\bibitem[{{Smith} {et~al.}(2010){Smith}, {Chornock}, {Silverman}, {Filippenko},
  \& {Foley}}]{smith2010}
{Smith}, N., {Chornock}, R., {Silverman}, J.~M., {Filippenko}, A.~V., \&
  {Foley}, R.~J. 2010, \apj, 709, 856, \dodoi{10.1088/0004-637X/709/2/856}

\bibitem[{{Spergel} {et~al.}(2013){Spergel}, {Gehrels}, {Breckinridge},
  {Donahue}, {Dressler}, {Gaudi}, {Greene}, {Guyon}, {Hirata}, {Kalirai},
  {Kasdin}, {Moos}, {Perlmutter}, {Postman}, {Rauscher}, {Rhodes}, {Wang},
  {Weinberg}, {Centrella}, {Traub}, {Baltay}, {Colbert}, {Bennett},
  {Kiessling}, {Macintosh}, {Merten}, {Mortonson}, {Penny}, {Rozo},
  {Savransky}, {Stapelfeldt}, {Zu}, {Baker}, {Cheng}, {Content}, {Dooley},
  {Foote}, {Goullioud}, {Grady}, {Jackson}, {Kruk}, {Levine}, {Melton},
  {Peddie}, {Ruffa}, \& {Shaklan}}]{spergel2013}
{Spergel}, D., {Gehrels}, N., {Breckinridge}, J., {et~al.} 2013, arXiv
  e-prints, arXiv:1305.5425.
\newblock \doarXiv{1305.5425}

\bibitem[{{Spergel} {et~al.}(2015){Spergel}, {Gehrels}, {Baltay}, {Bennett},
  {Breckinridge}, {Donahue}, {Dressler}, {Gaudi}, {Greene}, {Guyon}, {Hirata},
  {Kalirai}, {Kasdin}, {Macintosh}, {Moos}, {Perlmutter}, {Postman},
  {Rauscher}, {Rhodes}, {Wang}, {Weinberg}, {Benford}, {Hudson}, {Jeong},
  {Mellier}, {Traub}, {Yamada}, {Capak}, {Colbert}, {Masters}, {Penny},
  {Savransky}, {Stern}, {Zimmerman}, {Barry}, {Bartusek}, {Carpenter}, {Cheng},
  {Content}, {Dekens}, {Demers}, {Grady}, {Jackson}, {Kuan}, {Kruk}, {Melton},
  {Nemati}, {Parvin}, {Poberezhskiy}, {Peddie}, {Ruffa}, {Wallace}, {Whipple},
  {Wollack}, \& {Zhao}}]{spergel2015}
{Spergel}, D., {Gehrels}, N., {Baltay}, C., {et~al.} 2015, arXiv e-prints,
  arXiv:1503.03757.
\newblock \doarXiv{1503.03757}

\bibitem[{{Stanway} {et~al.}(2016){Stanway}, {Eldridge}, \&
  {Becker}}]{stanway2016}
{Stanway}, E.~R., {Eldridge}, J.~J., \& {Becker}, G.~D. 2016, \mnras, 456, 485,
  \dodoi{10.1093/mnras/stv2661}

\bibitem[{{Tanaka} {et~al.}(2013){Tanaka}, {Moriya}, \&
  {Yoshida}}]{tanaka2013beyond6}
{Tanaka}, M., {Moriya}, T.~J., \& {Yoshida}, N. 2013, \mnras, 435, 2483,
  \dodoi{10.1093/mnras/stt1469}

\bibitem[{{Terreran} {et~al.}(2017){Terreran}, {Pumo}, {Chen}, {Moriya},
  {Taddia}, {Dessart}, {Zampieri}, {Smartt}, {Benetti}, {Inserra},
  {Cappellaro}, {Nicholl}, {Fraser}, {Wyrzykowski}, {Udalski}, {Howell},
  {McCully}, {Valenti}, {Dimitriadis}, {Maguire}, {Sullivan}, {Smith}, {Yaron},
  {Young}, {Anderson}, {Della Valle}, {Elias-Rosa}, {Gal-Yam}, {Jerkstrand},
  {Kankare}, {Pastorello}, {Sollerman}, {Turatto}, {Kostrzewa-Rutkowska},
  {Koz{\l}owski}, {Mr{\'o}z}, {Pawlak}, {Pietrukowicz}, {Poleski}, {Skowron},
  {Skowron}, {Soszy{\'n}ski}, {Szyma{\'n}ski}, \& {Ulaczyk}}]{terreran2017pisn}
{Terreran}, G., {Pumo}, M.~L., {Chen}, T.~W., {et~al.} 2017, Nature Astronomy,
  1, 713, \dodoi{10.1038/s41550-017-0228-8}

\bibitem[{{Vink} \& {de Koter}(2005)}]{vink2005}
{Vink}, J.~S., \& {de Koter}, A. 2005, \aap, 442, 587,
  \dodoi{10.1051/0004-6361:20052862}

\bibitem[{{Vink} {et~al.}(2001){Vink}, {de Koter}, \& {Lamers}}]{vink2001}
{Vink}, J.~S., {de Koter}, A., \& {Lamers}, H.~J.~G.~L.~M. 2001, \aap, 369,
  574, \dodoi{10.1051/0004-6361:20010127}

\bibitem[{{Vreeswijk} {et~al.}(2014){Vreeswijk}, {Savaglio}, {Gal-Yam}, {De
  Cia}, {Quimby}, {Sullivan}, {Cenko}, {Perley}, {Filippenko}, {Clubb},
  {Taddia}, {Sollerman}, {Leloudas}, {Arcavi}, {Rubin}, {Kasliwal}, {Cao},
  {Yaron}, {Tal}, {Ofek}, {Capone}, {Kutyrev}, {Toy}, {Nugent}, {Laher},
  {Surace}, \& {Kulkarni}}]{vreeswijk2014}
{Vreeswijk}, P.~M., {Savaglio}, S., {Gal-Yam}, A., {et~al.} 2014, \apj, 797,
  24, \dodoi{10.1088/0004-637X/797/1/24}

\bibitem[{{Whalen} {et~al.}(2013){Whalen}, {Even}, {Frey}, {Smidt}, {Johnson},
  {Lovekin}, {Fryer}, {Stiavelli}, {Holz}, {Heger}, {Woosley}, \&
  {Hungerford}}]{whalen2013pisnzero}
{Whalen}, D.~J., {Even}, W., {Frey}, L.~H., {et~al.} 2013, \apj, 777, 110,
  \dodoi{10.1088/0004-637X/777/2/110}

\bibitem[{{Wiseman} {et~al.}(2020){Wiseman}, {Smith}, {Childress}, {Kelsey},
  {M{\"o}ller}, {Gupta}, {Swann}, {Angus}, {Brout}, {Davis}, {Foley},
  {Frohmaier}, {Galbany}, {Guti{\'e}rrez}, {Inserra}, {Kessler}, {Lewis},
  {Lidman}, {Macaulay}, {Nichol}, {Pursiainen}, {Sako}, {Scolnic}, {Sommer},
  {Sullivan}, {Tucker}, {Abbott}, {Aguena}, {Allam}, {Avila}, {Bertin},
  {Brooks}, {Buckley-Geer}, {Burke}, {Carnero Rosell}, {Carollo}, {Carrasco
  Kind}, {da Costa}, {De Vicente}, {Desai}, {Diehl}, {Doel}, {Eifler},
  {Everett}, {Fosalba}, {Frieman}, {Garc{\'\i}a-Bellido}, {Gaztanaga},
  {Gerdes}, {Gill}, {Glazebrook}, {Gruendl}, {Gschwend}, {Hartley}, {Hinton},
  {Hollowood}, {Honscheid}, {James}, {Kuehn}, {Kuropatkin}, {Lima}, {Maia},
  {March}, {Martini}, {Melchior}, {Menanteau}, {Miquel}, {Ogando},
  {Paz-Chinch{\'o}n}, {Plazas}, {Romer}, {Roodman}, {Sanchez}, {Scarpine},
  {Serrano}, {Suchyta}, {Swanson}, {Tarle}, {Thomas}, {Tucker}, {Varga},
  {Walker}, {Wilkinson}, \& {(DES Collaboration)}}]{wiseman2020}
{Wiseman}, P., {Smith}, M., {Childress}, M., {et~al.} 2020, \mnras, 495, 4040,
  \dodoi{10.1093/mnras/staa1302}

\bibitem[{{Wong} {et~al.}(2019){Wong}, {Moriya}, {Oguri}, {Hilbert}, {Koyama},
  \& {Nomoto}}]{wong2019ultimate}
{Wong}, K.~C., {Moriya}, T.~J., {Oguri}, M., {et~al.} 2019, \pasj, 71, 60,
  \dodoi{10.1093/pasj/psz037}

\bibitem[{{Woosley}(2010)}]{woosley2010}
{Woosley}, S.~E. 2010, \apjl, 719, L204, \dodoi{10.1088/2041-8205/719/2/L204}

\bibitem[{{Woosley} {et~al.}(2007){Woosley}, {Blinnikov}, \&
  {Heger}}]{woosley2007}
{Woosley}, S.~E., {Blinnikov}, S., \& {Heger}, A. 2007, \nat, 450, 390,
  \dodoi{10.1038/nature06333}

\bibitem[{{Yasuda} \& {Fukugita}(2010)}]{yasuda2010}
{Yasuda}, N., \& {Fukugita}, M. 2010, \aj, 139, 39,
  \dodoi{10.1088/0004-6256/139/1/39}

\bibitem[{{Yoon} {et~al.}(2012){Yoon}, {Dierks}, \& {Langer}}]{yoon2012}
{Yoon}, S.~C., {Dierks}, A., \& {Langer}, N. 2012, \aap, 542, A113,
  \dodoi{10.1051/0004-6361/201117769}

\bibitem[{{Yoon} {et~al.}(2010){Yoon}, {Woosley}, \& {Langer}}]{yoon2010}
{Yoon}, S.~C., {Woosley}, S.~E., \& {Langer}, N. 2010, \apj, 725, 940,
  \dodoi{10.1088/0004-637X/725/1/940}

\bibitem[{{Yoshida} {et~al.}(2014){Yoshida}, {Okita}, \& {Umeda}}]{yoshida2014}
{Yoshida}, T., {Okita}, S., \& {Umeda}, H. 2014, \mnras, 438, 3119,
  \dodoi{10.1093/mnras/stt2427}

\end{thebibliography}
\bibliographystyle{aasjournal}



\end{document}